\documentclass[fleqn,10pt]{wlscirep}
\usepackage[utf8]{inputenc}
\usepackage[T1]{fontenc}
\title{Social homophily predicts evacuation destination choice and long-term displacement decisions after the Marshall Fire}

\author[1,2]{Vaidehi Raipat}
\author[1,2,*]{Takahiro Yabe}
\affil[1]{Department of Technology Management and Innovation, Tandon School of Engineering, New York University, Brooklyn, NY 11201, USA}
\affil[2]{Center for Urban Science and Progress, Tandon School of Engineering, New York University, Brooklyn, NY 11201, USA}
\affil[*]{corresponding author: takahiroyabe@nyu.edu}

\begin{abstract}
Rapid urbanization and climate change have contributed to a significant rise in the frequency and intensity of disasters, which have resulted in three million adults being displaced from their homes in the United States during the past year, according to the Census Bureau.  
Using large-scale mobility data, it is now possible to observe and analyze post-disaster mobility dynamics across a longer time period, compared to household surveys which are often limited in scale. 
However, much of the mobility data-driven research on evacuation and displacement destination choice behavior has focused on analyzing the effects of physical and spatial factors, often underemphasizing the role of social factors, including individual preferences and social connections. 
Motivated by the recent literature showing strong associations between social homophily and mobility behavior, we use large-scale data of anonymized GPS traces and online social connections from the Marshall Fires in Colorado, USA, to unravel the associations between social and behavioral factors and evacuation destinations and long-term displacement decisions. 
We find that behavioral characteristics and social homophily play a significant role in post-disaster mobility decisions. 
First, incorporating pre-disaster behavior characteristics increases the predictability of evacuated distance by more than five-fold, which is critical for wildfire evacuation. 
Second, given the same evacuation distance, evacuees chose locations that have a high sociodemographic homophily with their home locations, and locations that have more friendship connections, compared to the level of homophily they are spatially exposed to. The social homophily effect is stronger among White and educated populations than low income, and Black and Asian populations. 
This effect has significant implications for long-term disaster impacts and policymaking, as it is a significant predictor of displacement and return decisions. 
Our findings highlight the importance of incorporating the social, behavioral, and economic characteristics to better understand and predict evacuation and displacement dynamics after disasters.
\end{abstract}
\begin{document}

\flushbottom
\maketitle
%
%
\thispagestyle{empty}

\section*{Introduction}

An increase in disasters driven by rapid urbanization, climate change, and other environmental factors has created an urgent need to better understand and predict the behavior patterns of affected individuals. These unprecedented environmental changes have also increased the incidence of wildfires throughout the country, causing property destruction and affecting large populations \cite{mcconnell2024rare, forrister2022evacuation}.
The movement patterns of disaster-affected populations typically unfold in distinct phases, including evacuation, return, and long-term displacement, each representing complex processes influenced by a variety of social, environmental, economic, and logistical factors \cite{martin2020using}.
A detailed understanding of these movement dynamics is crucial for designing resilient urban policies that enable effective disaster recovery, such as strategic prepositioning of resources and equitable allocation of post-disaster support to impacted communities \cite{aldrich2022social}.
From a behavioral science perspective, disaster evacuation and displacement patterns -- particularly following no-notice events such as wildfires -- offer a valuable natural experiment for examining the decision-making process of individuals. These scenarios enable us to unravel the critical factors that people prioritize when making mobility decisions \cite{lee2022specifying, martin2020using}. 

Disaster evacuation and displacement has been a long-studied topic in multiple fields, including civil and transportation engineering, social science, psychology, and complexity science \cite{thompson2017evacuation}. 
Household surveys \cite{aldrich2022social} and stated preference data have often been used to identify important factors for evacuation destination choice, including demographics, access to opportunities, existence of friends and family \cite{aldrich2012building,sadri2017modeling,lindell1992behavioral, forrister2022evacuation}, transportation convenience, distance between origin and destination \cite{jiang2021social, scheffer2020critical, stouffer1940intervening}.
More recently, studies using social media and mobile phone GPS location data (`mobility data') have reinforced such findings with more granular and longitudinal data of human behavior \cite{yabe2022mobile, lee2022specifying, naushirvanov2024evacuation}. Studies have revealed that factors such as community median income, population, housing damage \cite{mcconnell2024rare}, race \cite{deng2021high}, and education explain the heterogeneity in the initial and long-term displacement rates between communities \cite{yabe2020understanding}. 
However, due to the challenges of collecting large-scale social network and personal preference data, the social and behavioral dimensions of evacuation and displacement decisions have remained underexplored in mobility data-driven research.

Contrary to expectations that disasters disrupt and substantially alter mobility patterns, analysis of displacement after the Haiti Earthquake \cite{lu2012predictability} and various disaster events across the world \cite{wang2016patterns} reveal that post-disaster mobility closely mirrors pre-disaster behavior and is highly predictable, underscoring the stability of movement even amid crises. 
This raises the question of whether social and behavioral characteristics, which have been revealed as strong predictors of mobility patterns during non-disaster situations, also apply to post-disaster mobility dynamics.
For instance, studies leveraging geotagged online social media data demonstrate that friendship networks serve as strong predictors of mobility, emphasizing the critical role of social connections in shaping movement decisions \cite{cho2011friendship,chen2022contrasting}. 
Similarly, research on urban encounters reveals a strong tendency for social homophily in mobility dynamics, often described as `experienced segregation' \cite{nilforoshan2023human,moro2021mobility,yabe2023behavioral}. 
Together, these findings -- similarity between pre- and post-disaster mobility, and strong evidence of social effects on mobility behavior -- motivate us to explore the role of social networks and their qualities, such as homophily, in influencing evacuation and displacement movement decisions.


To this end, this study contributes to bridging this knowledge gap by leveraging large-scale, longitudinal, and anonymous mobile phone location data collected in the aftermath of a major disaster \cite{yabe2022mobile}. The Marshall Fire serves as a case study to test our hypotheses on post-disaster behavior. The Marshall Fire, a significant disaster in Colorado, began on December 30, 2021 as a forest fire and quickly spread into urban suburbs including Boulder, Louisville, and Superior. Over the course of three days, it displaced more than 3,000 residents and burned approximately 6,080 acres  \cite{noaa_marshall_fire, forrister2022evacuation, brulliard2022marshall}.
Overall, we find that both individual preferences and social connections significantly contribute to evacuation decisions and destination choice \cite{martin2020using, aldrich2022social}. Compared to the simulated null model of evacuation behavior, we show that individuals choose more socially similar and connected regions as evacuation destinations. Further analysis shows that this tendency varies across sociodemographic groups \cite{fussell2010race}, and that it has significant associations with long-term outcomes such as displacement and return \cite{fussell2010race, martin2020using}.  
These insights could inform urban policies for resource allocation and disaster management both in the short- and long-term after disasters. Such predictions could also help reinforce and prepare infrastructure systems in affected regions and address immediate needs in destination areas in the event of an unprecedented influx \cite{forrister2022evacuation}. 

\section*{Results}

Using a large and longitudinal dataset of anonymized individual GPS location records in Colorado from November 2021 to June 2022, we analyze the mobility patterns of individuals before, during, and after the Marshall Fire which occurred on December 30, 2021 to January 2, 2022. 
Anonymous, privacy-enhanced, and high-resolution mobile location ping data (`mobility data') of more than 200,000 devices in the State of Colorado was provided by Cuebiq. All devices within the study opted-in to anonymous data collection for research purposes under the General Data Protection Regulation (GDPR) and the California Consumer Privacy Act (CCPA) compliant framework. 
For privacy purposes, Cuebiq obfuscates device home locations at the census block group level (CBG) by estimating night time stay locations during a period of four weeks before the disaster at night time (8pm to 7am). Supplementary Note 1.2 \& 1.3, and Supplementary Figures S1 show that the mobility data samples have a high correlation ($R=0.821$, $p<0.001$) with the census population data, indicating that the data are representative of the population. Moreover, Supplementary Figure S2 \& S3 shows that the sample rate has a negligible correlation with household median income and race proportions in the granularity of CBGs, further indicating that the sample is well balanced. 

To detect evacuation behavior, a similar approach was employed to estimate the night time stay locations during and after the disaster. Each week, the CBG where each individual spent the most time during the night time period was labeled as the night time stay location for the week. An individual was labeled as evacuated if they spent the most time (and more than 480 minutes) in a CBG other than their home CBG during night time in the three days following the disaster. Supplementary Note 2.1 and Supplementary Figures S4 show the robustness of the results on evacuation behavior modeling against the selection of the minimum time threshold. To enable a longitudinal analysis, we include individuals whose data is continuously observed and are not missing for at most three consecutive weeks. To focus our analysis on the impacts of the fire, we focus on individuals whose home location was within a 100-kilometer radius of the epicenter.

\subsection*{Social determinants of evacuation behavior}
\begin{figure}[!h]
    \centering
    \includegraphics[width=0.85\linewidth]{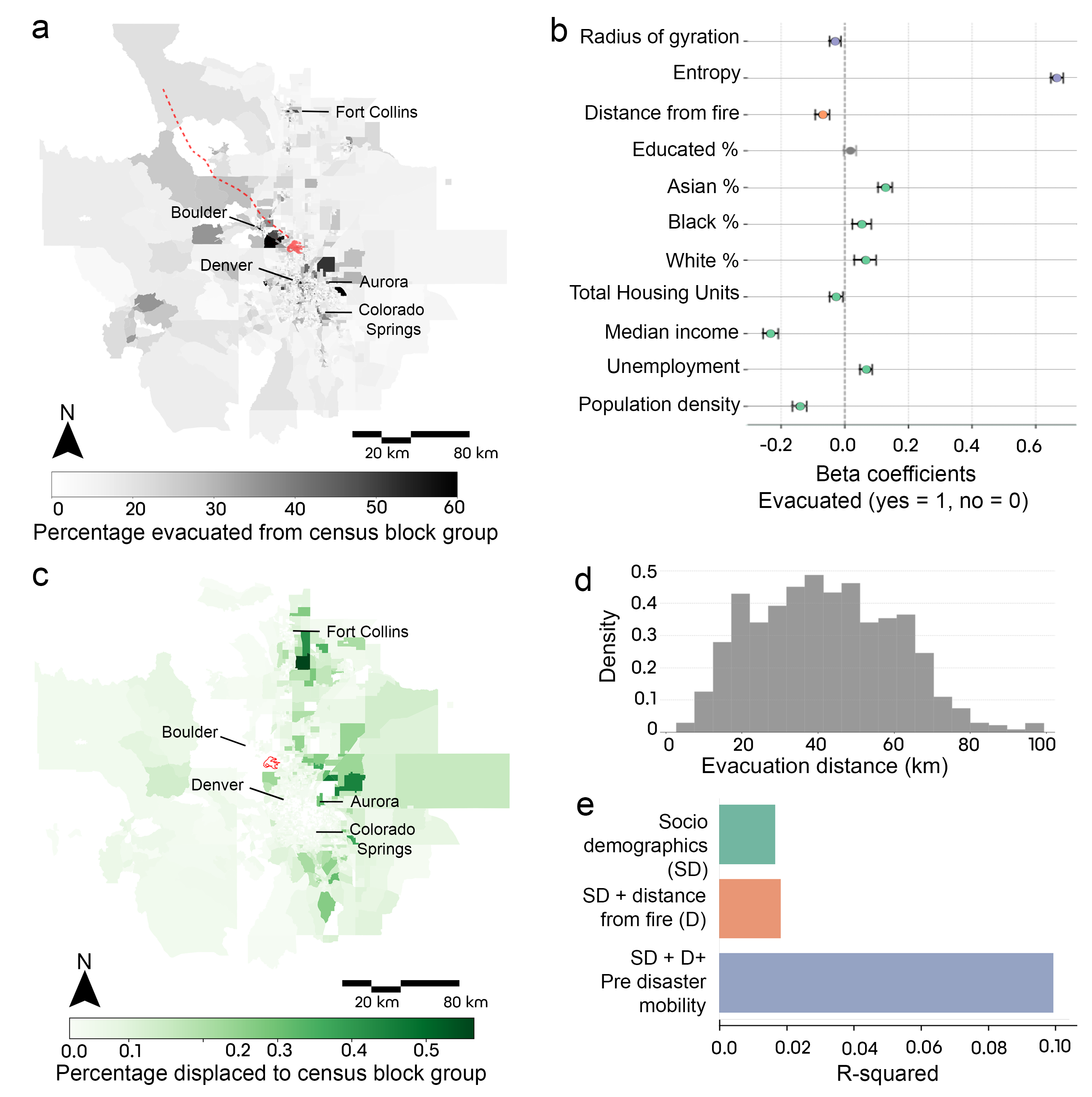}
    \caption{{\textbf{Determinants of evacuation decisions after Marshall Fire.} a) Spatial distribution of the percentage of people that evacuate from each census block group. Higher rates of evacuation were observed near the epicenter of the fire (red polygon) and along the path of the fire (red dotted line). b) Logistic regression coefficients for predicting individuals’ evacuation behavior. Significant and positive coefficient for entropy suggests that individuals having exploratory pre-disaster mobility behavior is the strongest predictor for evacuation behavior. The error bars show 95\% confidence intervals. c, d) Spatial distribution of evacuation destination shows that evacuees relocate to areas 20-60 km from the fire's epicenter (red polygon), indicating a preference for staying relatively close to home.  e) R-squared of each prediction model shows that pre-disaster mobility characteristics increase the predictability from 1.7\% to 9.8\%.
    Results suggest that individuals with higher radius of gyration before the disaster tend to travel further distances for evacuation. Maps were produced in Python using the TIGER shapefiles from the U.S. Census Bureau \cite{tiger}.
 }}
    \label{fig:Figure 1}
\end{figure}

Using the weekly night time stay locations of each individual before, during, and after the fire, we first analyze the social and spatial determinants of evacuation dynamics. 
Figure \ref{fig:Figure 1}a shows the evacuation rates, defined as the proportion of individuals in the data who evacuated in the first 3 days after the fire, in census block groups (CBGs) within 100 kilometers from the epicenter of the fire. 
As expected, we observe a strong spatial pattern of evacuation behavior, where CBGs along the path of the fire (red dotted line) or very close to the fire (red polygon) have higher evacuation rates. 
To understand the factors associated with evacuation decisions, we tested a logistic regression model with the specification: $Pr(y_u = 1) \sim logit^{-1}[G_u+SD_u+B_u]$, where $y_u=1$ indicates the user $u$ evacuated and $y_u=0$ otherwise. Independent variables include $G_u$, $SD_u$, and $B_u$, which are groups of geographical, sociodemographic, and behavioral characteristics, respectively. 
We find that, in addition to the distance from the individual's home to the epicenter of the fire, and the sociodemographic characteristics of the individual's home location CBG, pre-disaster mobility behavior patterns played a significant role in determining their evacuation decisions, as shown in Figure \ref{fig:Figure 1}b. 
More specifically, metrics such as the individual's mobility entropy $H$ ($\beta = 0.834$, p-value $<0.001$), which measures the diversity of visits across different locations based on their total dwell time in each location, and the radius of gyration $R_g$ ($\beta = -0.0357$, p-value $<0.001$), which measures the typical distance a user travels based on their visited locations and the time spent at each location, were strong predictors of evacuation decisions (see Methods for more details on the pre-disaster mobility metrics). The full regression tables are shown in Supplementary Table S1.

For effective planning and pre-positioning of resources, understanding how individuals choose evacuation destinations is of substantial importance. Figure \ref{fig:Figure 1}c maps the percentages of individuals evacuated to each census block group. We find that, by plotting the histogram of evacuation distances in Figure \ref{fig:Figure 1}d, that the majority of evacuees (64.76\%) move to areas between 20 and 60 km from the epicenter of the fire, indicating a preference to stay relatively close to their home location. To understand the factors associated with the individuals' evacuation distances $r_u$, we tested a simple linear regression model with the specification: $log_{10}(r_u) \sim G_u+SD_u+B_u$, where $r_u$ indicates the user $u$'s evacuation distance, and $G_u$, $SD_u$, and $B_u$, are groups of geographical, sociodemographic, and behavioral characteristics, respectively, similar to the previous logistic regression in Figure \ref{fig:Figure 1}b. 
By comparing the incremental increase in R-squared with the inclusion of independent variable groups $G_u$, $SD_u$, and $B_u$, we find that the inclusion of pre-disaster mobility behavior $B_u$ substantially increases the model’s predictability, by more than five-fold (from $R^2= 0.017$ to $R^2= 0.098$) (Figure \ref{fig:Figure 1}e). The coefficient of the radius of gyration is highly significant and positive ($\beta = 0.2086$, p-value $<0.001$), suggesting that evacuees with a higher radius of gyration tend to travel further distances when evacuating. The full regression tables are shown in Supplementary Tables S6, S7 \& S8. 
These results suggest that the behavioral factors of individuals are strong determinants of evacuation decisions.

\subsection*{Social homophily between home and evacuation destinations}

\begin{figure}[!h]
    \centering
    \includegraphics[width=0.9\linewidth]{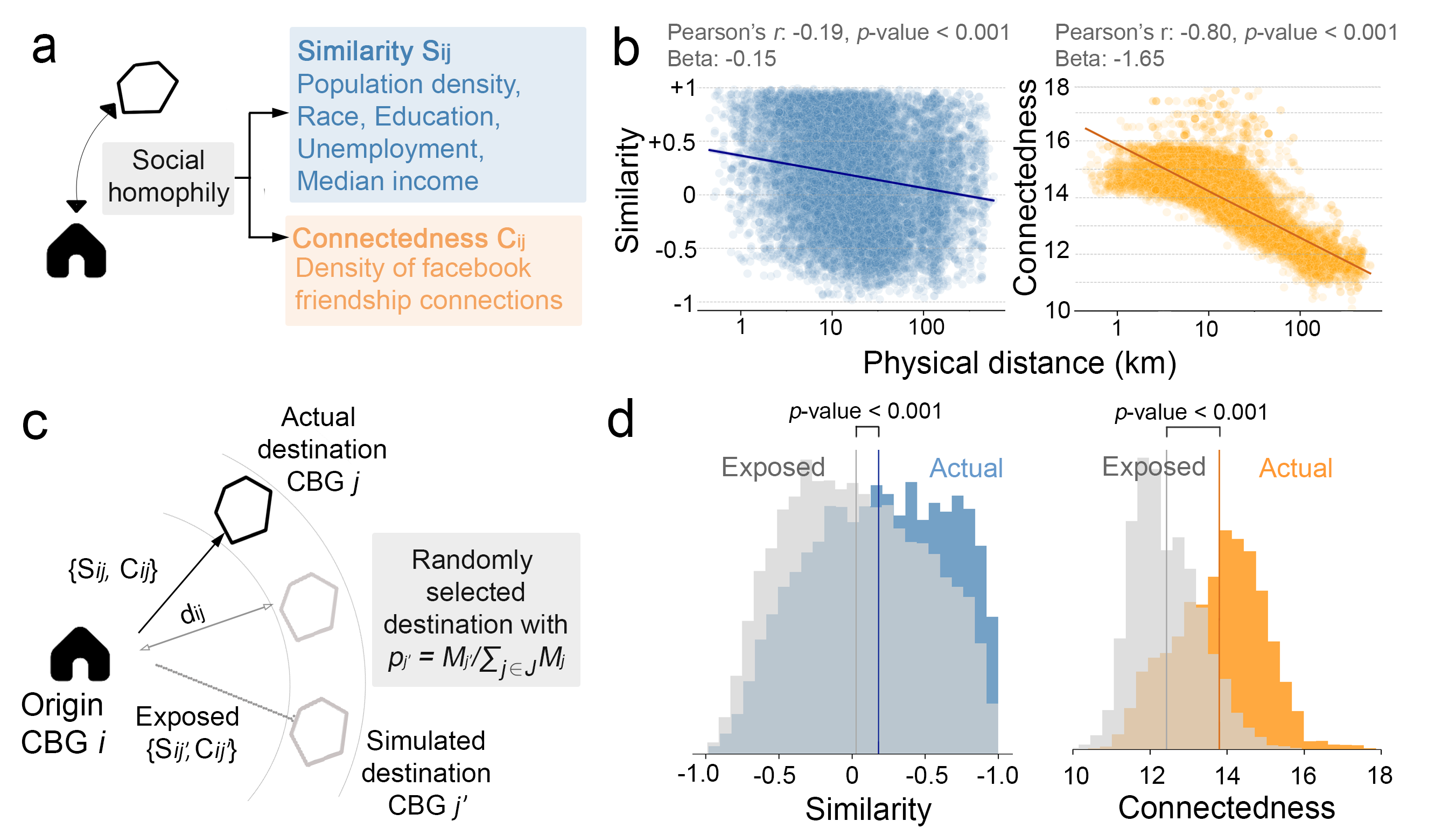}
    \caption{{\textbf{Social homophily and destination choice} a) Two metrics for social homophily are studied: Similarity $S_{ij}$ is defined as the cosine similarity between origin ($i$) and destination ($j$) census block groups, using demographic attributes including median income, population density, racial composition, and percentage of educated population. Connectedness $C_{ij}$ between $i$ and $j$ is defined by the number of friendship connections between these locations as represented on Facebook. b) Similarity and connectedness decay with distance between $i$ and $j$, indicating that individuals who evacuate longer distance are more likely to stay in less similar and connected neighborhoods.  c) The null model for evacuation destination simulation. For each individual, the destination is randomly selected proportional to the probability $p_{ij} = M_j/d_{ij}^2$ derived from the gravity model. This null model assumes that individuals choose evacuation destinations based on distance and population density, ignoring social factors. d) Similarity $S_{ij}$ (blue box plots) and connectedness $C_{ij}$ (orange box plots) are calculated for both actual and simulated destinations. Results show higher similarity and connectedness for actual destinations compared to the null model, indicating that social homophily substantially affects the choice of destinations during evacuation. Figures 2a and 2c were designed using icons from the Noun Project created by pictranoosa. 
}}
    \label{fig:Figure 2}
\end{figure}

The results on evacuation behavior so far indicate that the evacuation behavior of individuals is strongly associated with the individuals' pre-disaster behavioral traits. 
Similarly, studies suggest that the individuals' preferences and social ties are important in evacuation destination choice \cite{fussell2010race, yabe2022mobile, aldrich2022social, collins2018effects, czaika2022migration}.
Here, we leverage the high granularity of the mobility data to assess the effects of social homophily on destination choice. More specifically, we analyze the effect of social homophily on evacuation destination choice by measuring i) sociodemographic similarity (``Social Similarity'' $S_{ij}$) and ii) the number of friendship connections (``Social Connectedness'' $C_{ij}$) between the destination $j$ and the home location (origin) $i$, as shown in Figure \ref{fig:Figure 2}a. The similarity index $S_{ij}$ is computed as the cosine similarity between the vectors of sociodemographic characteristics of the origin and destination CBGs. Sociodemographic characteristics include population density, race proportions, level of education, unemployment rates, and the median income of the CBG. Supplementary Note 3.1 - 3.4 contains the robustness checks on the selection of the sociodemographic characteristics. The social connectedness $C_{ij}$ is measured using Facebook's Social Connectedness Index (SCI), which measures the probability of a friendship connection on Facebook between two locations (zipcode).  
Compared with the physical distance $d_{ij}$ between CBGs $i$ and $j$, we find that both the similarity ($\beta = -0.19$, $p<0.001$) and connectedness ($\beta = -0.80$, $p<0.001$) metrics decay with distance, as shown in Figure \ref{fig:Figure 2}b. This suggests that individuals who evacuate longer distances are more likely to stay in less similar and connected neighborhoods, which could lead to fewer social connections and support, which are suggested in the literature to be essential ingredients for effective disaster response and recovery \cite{aldrich2012building}.

Studies have used the gravity model as the standard model to characterize and model evacuation patterns after disasters \cite{cheng2011dynamic,jiang2021social}. However, behavioral preferences and the characteristics of the destination could also play a crucial role in evacuation patterns. Despite the availability of large scale mobility datasets, such dynamics have been rarely quantified. 
To investigate the significance of behavioral preferences, we construct a null model to simulate evacuation destinations for each individual \cite{zhang2024counterfactual}. In the null model, illustrated in Figure \ref{fig:Figure 2}c, counterfactual destinations $j'$ are randomly simulated for each individual with probabilities proportional to the gravity model $p_{ij} = M_j/d_{ij}^2$, which ignores any social and behavioral preferences. 
The randomly simulated counterfactual destinations $j'$ consider the spatial configurations of surrounding cities and communities and their sizes, allowing us to quantify the level of social similarity and connectedness to which individuals are spatially exposed.
Figure \ref{fig:Figure 2}d shows the distribution of social similarity $S_{ij'}$ and social connectedness $C_{ij'}$ for actual and simulated destinations. The distribution of the similarity and connectedness measures of the actual destinations is skewed towards higher values, compared to the simulated destinations using the null model. The gray vertical lines in both figures show the mean values of the simulated destinations, and the blue and orange vertical lines show the mean values of actual destinations. The differences in the means are statistically significant in both cases with a $p < 0.001$, indicating that social homophily is an important factor in evacuation destination choice decision, and that leveraging a simple gravity model could overlook the complexity of decision making during disaster evacuation. 

\subsection*{Demographic inequity in social homophily of destinations}

\begin{figure}[!h]
    \centering
    \includegraphics[width=0.85\linewidth]{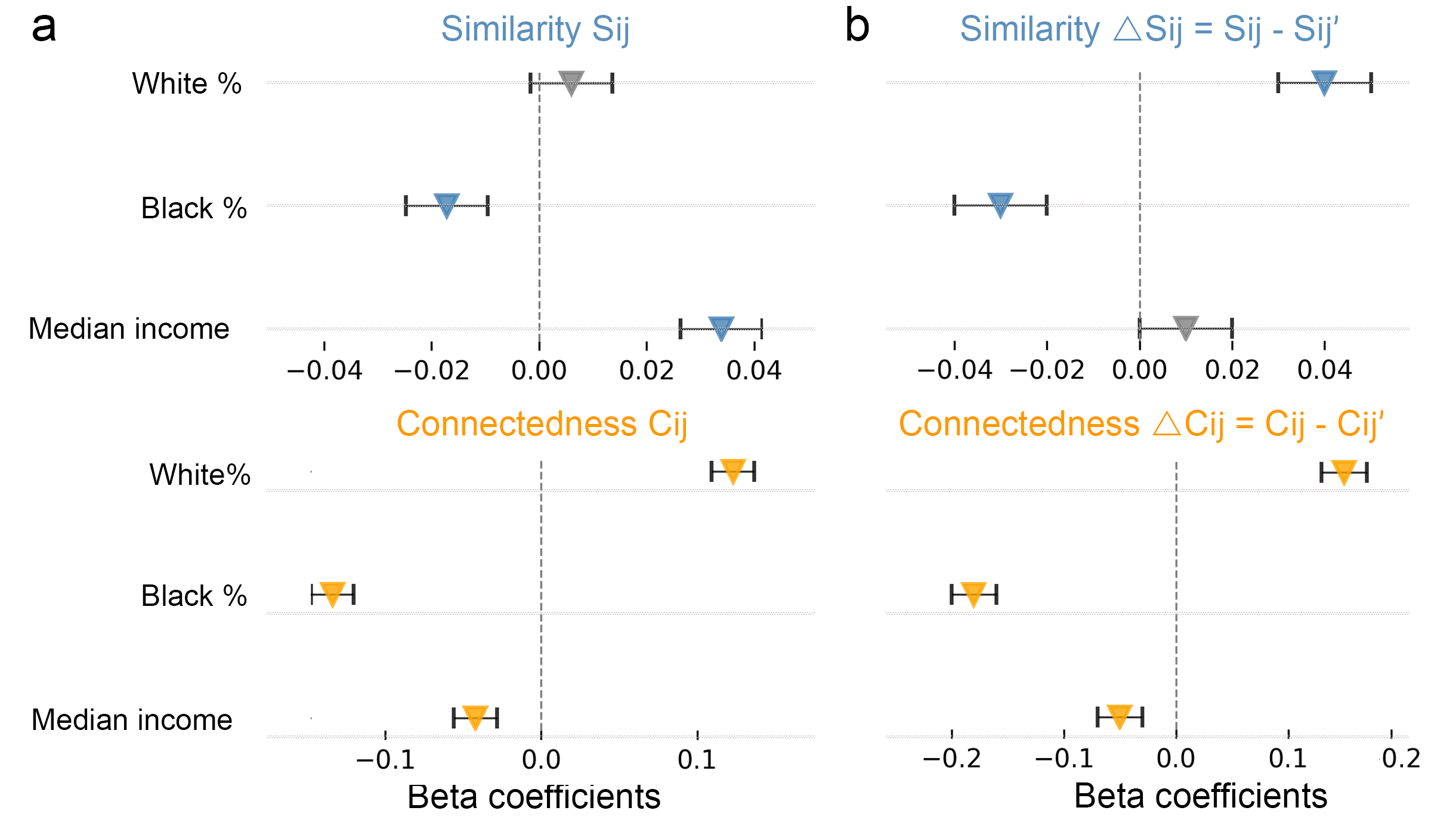}
    \caption{{\textbf{Sociodemographic inequity in social homophily after evacuation} a) Regression coefficients for $\{S_{ij}, C_{ij}\} \sim d_{ij} + {SD_i}$, where ${SD_i}$ are sociodemographic variables including White percent, Black percent, and median income of the origin census block group. Controlling for evacuation distance ${d_{ij}}$, we find that white and median-income populations are more likely to evacuate to destinations that have high social homophily with their origin. b) Regression coefficients for $\{\Delta S_{ij}, \Delta F_{ij}\} \sim d_{ij} + {SD_i}$, where ${SD_i}$ are sociodemographic variables including White percent, Black percent, and median income of the origin census block group. Evacuees from neighborhoods with higher proportions of White residents chose destinations significantly more similar and connected compared to simulated destinations that were based on distance and population density only, indicating the significant effects of behavioral preferences. In both plots, the error bars show 95\% confidence intervals. 
}}
    \label{fig:Figure 3}
\end{figure}

The counterfactual simulations developed in Figure \ref{fig:Figure 2}c have shown that evacuees tend to prefer destinations that have higher sociodemographic similarity and social connections, conditioned on the spatial configuration of cities and their sizes. However, it is unclear whether this tendency is homogeneous between different sociodemographic groups \cite{sun2024social}.
The potential inequity between demographic groups is important to quantify, due to the possible difficulties individuals from more vulnerable populations may face when evacuating to areas that are unfamiliar or communities lacking social ties \cite{deng2021high, sun2024social}. 
To measure this heterogeneity, we use linear regression to model the association between the evacuees' demographic characteristics ${SD_i}$ and the social homophily of realized destinations ${S_{ij}, C_{ij}}$, controlled on the evacuation distance $d_{ij}$. Thus, the regression model is: $\{ {S_{ij}, C_{ij}} \} \sim d_{ij} + {SD_i}$. 
As shown in Figure \ref{fig:Figure 3}a, the regression coefficients for White percentage ($\beta = 0.123$, $p< 0.001$) and median household income ($\beta = 0.033$, $p< 0.001$) are positive and significant, indicating that evacuees who live in census block groups (CBG) with higher median household income and higher percentage of white population have a greater tendency to select connected and similar destinations, respectively. On the other hand, individuals with higher proportions of Black populations had a higher tendency to evacuate to areas with lower levels of sociodemographic similarity and social connectivity. 
The full regression tables are shown in Supplementary Table S9 to S14.

The observed inequity in realized social homophily during evacuation, shown in Figure \ref{fig:Figure 3}a, is a significant and substantial difference, however, could be due to the lack of options for Black populations (i.e., there are fewer communities that are sociodemographically similar and socially connected with such communities). 
To test whether this inequity is the result of behavioral preferences or due to the lack of opportunities, we leverage the simulated evacuation patterns from Figure \ref{fig:Figure 2}. 
For each individual, we compute the difference in similarity $\Delta S_{ij} = S_{ij} - S_{ij'}$ and connectedness $\Delta C_{ij} = C_{ij} - C_{ij'}$. We regress the differences in similarity and connectedness between the actual and simulated destinations using a linear regression: $\{ \Delta S_{ij}, \Delta C_{ij} \} \sim d_{ij} + {SD_i}$, where ${SD_i}$ are sociodemographic variables including White percentage, Black percentage, and the median income of the origin census block group. 
The results, shown in Figure \ref{fig:Figure 3}b, reveal significant and positive coefficients for the White population percentage, indicating that evacuees from predominantly White CBGs selected destinations that were more similar and connected than predicted by the gravity-based model. These results advocate for the incorporation of behavioral preferences in predictive evacuation models, enhancing their ability to accurately reflect observed patterns of movement and providing a more robust framework for disaster response planning. Supplementary Tables S15 to S20 show the full regression results.

\subsection*{Social homophily during evacuation is associated with long-term displacement outcomes}
\begin{figure}[!h]
    \centering
    \includegraphics[width=0.85\linewidth]{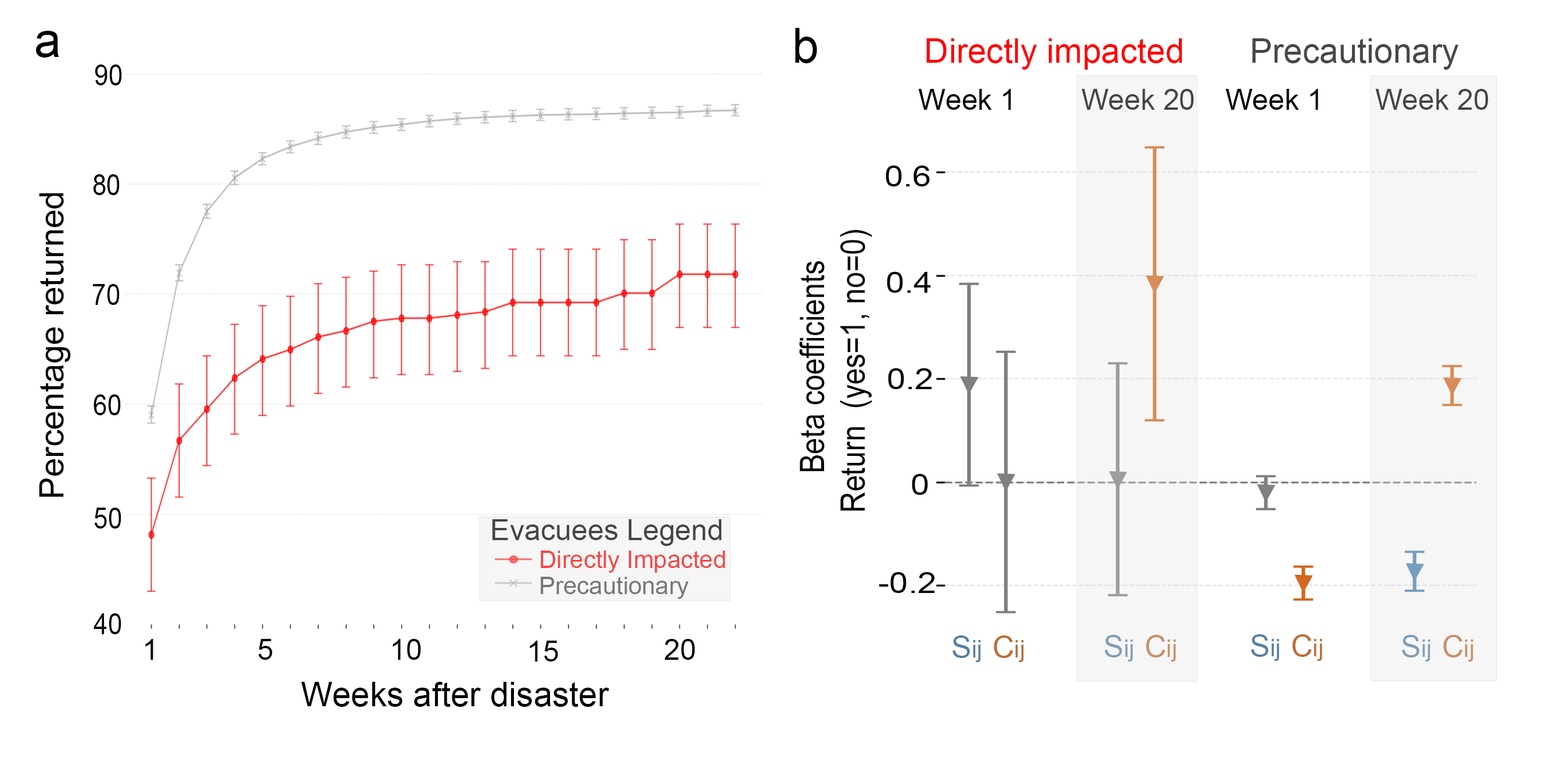}
    \caption{{\textbf{Social homophily during evacuation is associated with long-term displacement outcomes.} a) Weekly percentage of evacuees who returned after the fire. The return rate is consistently higher for evacuees located in the burned regions compared to those outside. 84.1 percent of evacuees from the non-burned areas and 64.1 percent of evacuees from the burned areas returned by the 8th week. The recovery rate slows significantly after week 8. The error bars show 95\% confidence intervals estimated using bootstrapping method. 
    b) Regression coefficients of long-term displacement outcomes (1=return, 0=displaced) on the similarity and connectedness of destination locations. The similarity and connectedness of evacuation destinations in the long-term (22 weeks after the fire) had significant impacts on their displacement outcomes. Evacuees who chose connected destinations returned more ($\beta = 0.186$, $p<0.001$) in the long-term (20 weeks after the fire). Choosing a similar destination location had the opposite effect for evacuees from non-burned areas, decreasing the likelihood of returning to their original home locations ($\beta = -0.173$, $p<0.001$). The error bars show 95\% confidence intervals. 
}}
    \label{fig:Figure 4}
\end{figure}

Understanding and predicting long-term patterns of individual movements, including displacement and return \cite{lee2022specifying}, is essential to ensure resilience and recovery of disaster-affected areas \cite{paul2024household, rivera2020impact}. The results in Figures \ref{fig:Figure 1} to \ref{fig:Figure 3} show the importance of utilizing behavioral preferences in modeling evacuation decisions and destination choice, however, how do these insights inform long-term outcomes? 
To answer this question, we investigate whether the social homophily of the evacuation destination has any association with the decision to return or not in the long term. Figure \ref{fig:Figure 4}a shows the percentage of individuals who returned to their original home location CBG after evacuation, across 22 weeks after the disaster. 
The red curve shows the return percentage of individuals who were directly impacted (home was located in the CBGs that were recorded as burnt) and evacuated. Their return rate is slower and lower (66.7 percent of directly impacted evacuees return by week 8 and 71.8 percent by week 22). 
The gray curve shows the return percentage of those who were living outside the directly impacted zone (experienced the fire) and evacuated. These individuals have a shorter period until return and an overall higher return percentage (84.7 percent return by week 8 after which the recovery rate slows significantly). Bootstrapping was used to compute the confidence intervals (error bars show 95\% confidence intervals) for each week's return percentage of precautionary and directly impacted evacuees to show the statistical significance of the difference in return percentages. 

Figure \ref{fig:Figure 4}b shows the regression coefficients of similarity and connectedness of destination locations on long-term displacement outcomes. More specifically, we conducted a logistic regression with the specification: $Pr(r_u = 1) \sim logit^{-1}[S_{ij}+C_{ij}]$, where $r_u=1$ indicates that the user $u$ returned and $r_u=0$ indicates displacement. Independent variables $S_{ij}$ and $C_{ij}$ are the social similarity and social connectedness between the user's pre-disaster home CBG and evacuation destination CBG. The results suggest that similarity and connectedness of the evacuation destinations in the long-term (22 weeks after the fire) had significant impacts on their displacement outcomes. Evacuees who chose destinations with high social connectedness had higher rates of return ($\beta = 0.186$, $p<0.001$). Choosing a destination with high social similarity had the opposite effect for evacuees, decreasing their likelihood of returning to their original home locations ($\beta = -0.173$, $p<0.001$). The mixed results of social homophily on return and displacement provide nuanced insights on how we may be able to predict long-term post-disaster movement outcomes by assessing the social characteristics of where the individual evacuated to. The full regression tables are shown in Supplementary Tables S21 to S28.

\section*{Discussion}


Cities and regions worldwide are increasingly exposed to the risks of climate change-induced disasters. Understanding how individuals make short- and long-term decisions during evacuation, return, and displacement is critical to developing resilient urban policies \cite{halofsky2020changing}.
While extensive research has explored the relationships between social, demographic, geographic factors, hazard characteristics (e.g., seismic intensity), and evacuation behaviors, these studies often rely on household surveys \cite{aldrich2022social}. However, such surveys are frequently constrained by limited spatial and temporal coverage and issues of representativeness. The availability of passively collected, large-scale behavioral datasets, such as mobile phone GPS location data and social media data, has provided new opportunities to analyze evacuation dynamics \cite{yabe2022mobile}. Despite this progress, there has been limited effort to develop computational approaches that test hypotheses about the social and behavioral determinants of evacuation and displacement. Furthermore, there remains a critical gap in understanding how these characteristics influence long-term outcomes for affected populations.

In this context, this study makes three contributions to understanding the social \cite{collins2018effects} and behavioral determinants of evacuation and displacement dynamics, using large-scale mobile phone GPS location data collected from individuals affected by the Marshall Fire in Colorado. 
First, our analysis reveals that pre-disaster mobility characteristics are the strongest predictors of both the likelihood of evacuation and the distance traveled during evacuation (Figure \ref{fig:Figure 1}). 
Specifically, individuals with less predictable movement patterns before the disaster were more likely to evacuate, and those with a greater radius of gyration tended to travel farther distances. 
This suggests that individuals who typically travel farther distances may be more accustomed to managing unfamiliar environments, making them more likely to relocate further during crises. 
These findings align with and extend prior research on mobility profiles, such as the classification of explorers and returners \cite{pappalardo2015returners}, by showing that these characteristics also have implications on mobility behavior during emergencies. This finding also adds more nuance to the previous findings about how post-disaster mobility mirrors pre-disaster mobility \cite{lu2012predictability,wang2016patterns}. 
From a practical perspective, linking individual-level mobility patterns before and after the disaster can enhance our ability to anticipate evacuation behavior and identify populations that are more likely to evacuate to further locations.


Second, our analysis of evacuation destination choices based on social connectedness and similarity reveals that evacuees, when controlled for distance, tend to select destinations that have higher social homophily. White and educated populations had a stronger tendency to select destinations with high similarity, while lower-income and minority populations were less likely to choose connected destinations. These findings mirror the social homophily observed in non-disaster mobility patterns using Twitter data from US cities \cite{wang2018urban, deng2021high}.
The observed differences in the tendency for social homophily during evacuation between sociodemographic groups suggest potential disparities in access to social capital \cite{chetty2022social,aldrich2022social}. This could help prioritize resource allocation in improving access to shelters, especially for the vulnerable population who lack social capital.
Finally, evacuees who relocated to destinations with high social connectedness or demographic similarity exhibited slower return rates or, in some cases, did not return at all. This could reflect the influence of social ties and support systems that provide stability, making it less urgent or necessary for evacuees to return to their original homes and also less stressful in cases where their home faces tremendous damages \cite{sadri2018role}. The findings underline the role of homophily -- the preference for social interactions with similar others -- in shaping both immediate evacuation decisions and long-term recovery processes. 
This finding suggests the importance of disaster relief programs to consider the role of community bonds in supporting displaced individuals. It also reinforces the potential benefits of enabling evacuees to resettle in socially supportive environments. \cite{hawkins2010bonding, chetty2022social1}

Our study has several limitations. Although the GPS location data used in the study allow us to observe and analyze how social homophily is associated with evacuation patterns, we are not able to observe the actual social interactions (e.g., support and altruistic behavior) between the evacuees and the recipient population. Combining this analysis with data collected from household surveys, interviews, and focus groups in this context could enhance our understanding of the mechanism of how evacuating to an area with high social homophily affects the post-disaster recovery outcomes of individuals and households. 
The social connectedness index data provided by Meta captures the relationships between users on Facebook. Although numerous studies have used this dataset to analyze social-spatial network structures, this could overlook non-digital forms of social capital or networks outside of this platform. Integrating alternative data sources to triangulate social-spatial connectivity could further strengthen the insights obtained in this study. 

The findings on the social and behavioral determinants of evacuation, displacement, and return mobility decisions highlight several directions for future research. In this study, we analyzed post-disaster mobility patterns within the specific context of a single major disaster. While we do not claim or hypothesize that these findings are universally applicable across other disasters in different geographies and contexts, our results provide concrete evidence of the critical role that social and behavioral determinants play in shaping post-disaster mobility dynamics. Furthermore, this research offers a computational and analytical framework that can be applied to datasets collected from other disasters to explore similar questions. A valuable next step would be to classify disaster typologies based on whether social and behavioral determinants have positive or negative effects on mobility decisions. Such a comprehensive approach could deepen our understanding of the universality and variability of how different factors affect post-disaster mobility patterns, and inform more targeted and effect disaster resilience strategies. 


\section*{Methods}

\subsection*{Evacuation behavior detection}
We leverage a large and longitudinal dataset of anonymized individual GPS location mobility data provided by Cuebiq to analyze evacuation, destination choice, return and displacement patterns among individuals residing in Colorado, focusing on nighttime dwelling patterns pre- and post-disaster. We gather daily stop data within the specified period, organizing and processing it into weekly segments to identify nighttime location across census block groups (CBGs). For each day, we retain the top three CBGs where the user spends the most nighttime dwell time, enabling a high-resolution view of their primary recurring locations. If a user consistently has a maximum nighttime (8pm - 7am) dwell time (minimum of 480 minutes) in the same CBG for four weeks pre-disaster (November 08– December 02, 2021), this CBG is defined as their home location.
 
To detect evacuation behavior, individuals were classified as evacuated if they spent more than 480 minutes of nighttime at a census block group (CBG) that was not their home CBG during the three days of the disaster event. Similarly, post-disaster home locations were identified for individuals residing within 100 km of the disaster epicenter by analyzing nighttime dwell durations exceeding 480 minutes. The dataset spans four weeks pre-disaster, the disaster period (December 31, 2021–January 2, 2022), and twenty-two weeks post-disaster (January 3–June 6, 2022). To ensure robust location assignment, only individuals with consistently observed data allowing for a maximum of three consecutive weeks of missing data were included in the analysis. Nighttime dwell time exceeding the specified threshold of 480 minutes served as the primary criterion for determining both evacuation status and home locations.

\subsection*{Pre-disaster mobility behavior}
We use two metrics to analyze the pre-disaster mobility behavior of selected evacuees by calculating two metrics: entropy $H$ and radius of gyration $(R_g)$ . These are computed based on all available stops for each user recorded mobility data from November 1st to 30th, 2021 and capture the spatial and temporal patterns of an individuals movement across visited locations.
We use entropy $H$ to quantify the diversity (or unpredictability of) an individual's mobility.
Higher entropy values indicate visits spread more evenly across multiple locations, suggesting greater diversity and low predictability, while lower entropy values indicate more concentrated movement patterns and high predictability. 
Shannon entropy, $H = -\sum (p_k \log(p_k))$, where  $p_k$  represents the probability of visiting a specific location $k$ \cite{shannon1948mathematical}, is computed using the \texttt{scipy.stats.entropy} function in Scipy \cite{scipy_entropy}.
We define $p_k$ as the proportion of total dwell time spent at a location relative to the total dwell time across all locations visited by the user. Shannon entropy thereby provides a metric for assessing the variability or concentration of an individual’s location choices over time.
The radius of gyration ($R_g$) \cite{pappalardo2015returners} represents the typical distance a user travels on an average day, capturing spatial dispersion. The $R_g$ weighted by dwell time is computed as $R_g = \sqrt{\frac{1}{N} \sum_{i \in L} n_i (r_i - r_{cm})^2}$, where $L$ is the set of locations, $r_i$ the geographic coordinates of location $i$, $n_i$ the time spent at location $i$, and $r_{cm}$ the center of mass of all visited locations. This formula extends the standard radius of gyration by incorporating $n_i$ as a weight, reflecting the significance of frequently visited locations.

\subsection*{Evacuation and destination rates}

The evacuation rate for each census block group (CBG) was calculated as the proportion of the population within the CBG that evacuated following the disaster. 
Similarly, destination rate for each CBG was calculated to visualize the spatial distribution of evacuees’ chosen destinations and represents the proportion of evacuees who relocated to a specific CBG relative to the total evacuated population.
To understand the spatial distribution of evacuation rates, we calculated the proportion of individuals evacuating within 100 km of the fire. A logistic regression model was employed to analyze factors influencing evacuation decisions, with the specification: $\text{Pr}(y_u = 1) \sim \text{logit}^{-1}[G_u + SD_u + B_u]$ where $y_u = 1$ indicates evacuation, $G_u$ includes geographic characteristics, $SD_u$ includes sociodemographic characteristics, and $B_u$ includes behavioral characteristics. 
Similarly, to understand factors associated with evacuation distances ($r_u$), a linear regression model was specified:
$\log_{10}(r_u) \sim G_u + SD_u + B_u$
where $r_u$ is the evacuation distance, and $G_u$, $SD_u$, and $B_u$ represent geographic, sociodemographic, and behavioral characteristics, respectively.

\subsection*{Social homophily}

Homophily is a sociological concept that describes the tendency of individuals to associate and bond with similar others. \cite{Figueredo:2009dg, mcpherson2001birds} In the context of this analysis, we have measured homophily using two metrics: connectedness, which represents the social connections between two locations, and sociodemographic similarity, which represents the demographically similarity of the two locations. 
To calculate social similarity, we use various demographic variables such as race, income, education, and unemployment of the origin and destination. Cosine similarity is used to measure the similarity between the pair of CBGs. 
To calculate social connectedness, we use the Social Connectedness Index (SCI) provided by Meta \cite{bailey2018social}. This index measures the strength of social connections between two geographic areas in the granularity of the zipcode, as represented by Facebook friendships. They use an anonymized snapshot of all active Facebook users and their friendship networks to measure the intensity of connectedness between locations. SCI measures the relative probability of Facebook friendship links between two locations $i$ and $j$ because it is scaled from maximum value of 1,000,000,000 and a minimum value of 1 and a random noise is added to round the number to the nearest integer.

\subsection*{Counterfactual simulations of destinations}
To investigate the significance of behavioral preferences in evacuation destination choice, we construct a null model to simulate evacuation destinations for each individual. In the null model, each evacuee's counterfactual evacuation destination $j'$ is randomly simulated for each individual with probabilities proportional to the gravity model $p_{ij} = M_j/d_{ij}^2$, among a set of candidate CBGs that are located at a similar distance from the home location as the actual evacuation destination. This weighted sampling method reflects the idea that destinations with larger populations exert a greater pull on evacuation behavior.  Using these probabilities, one destination per individual was sampled. 
The connectedness and similarity levels between the actual and simulated destinations by calculating the mean similarity and familiarity for both. Linear regression was employed to examine the relationship between evacuees’ demographic characteristics ($SD_i$) and the social homophily of their chosen destinations ($S_{ij}, C_{ij}$), while controlling for evacuation distance ($d_{ij}$). Furthermore, the differences between actual and simulated destinations in social similarity and social connectedness $\{\Delta S_{ij}, \Delta C_{ij}\}$, was regressed to measure the roles of structural and behavioral factors in shaping evacuation choices.

\section*{Data Availability}
The data that support the findings of this study are available from Cuebiq through their Social Impact program, but restrictions apply to the availability of these data, which were used under the license for the current study and are therefore not publicly available. Information about how to request access to the data and its conditions and limitations can be found in \url{https://cuebiq.com/social-impact/}.  
Data access requests should be submitted through Cuebiq's Social Impact customer page \url{ https://cuebiq.com/demo/}, where the Sales team at Cuebiq may be contacted. Other data including the American Community Survey is available for download at \url{https://data.census.gov/}, and Tiger shapefiles can be downloaded from the US Census Bureau \url{https://www.census.gov/programs-surveys/geography/guidance/tiger-data-products-guide.html}. 

\section*{Code Availability}
The analysis was conducted using Python. Code to reproduce the main results in the figures from the aggregated data is publicly available on GitHub \url{https://github.com/takayabe0505/socialhomophily}.


\bibliography{sample}

\section*{Acknowledgements}
We would like to thank Spectus who kindly provided us with the mobility dataset for this research through their Data for Good program. T.Y. acknowledges support by the National Science Foundation under Grant number 2343646.

\section*{Author contributions statement}
All authors designed the algorithms, performed the analysis, developed models and simulations, and wrote the paper. Company data were processed by T.Y. and partially by V.R. T.Y. had access to aggregated (nonindividual) processed data. All authors reviewed the manuscript. 

\section*{Competing Interests Statement}
The authors declare no competing interests.

\end{document}


\maketitle

\tableofcontents
\newpage

\listoffigures

\listoftables

\newpage

\setcounter{figure}{0}
\setcounter{table}{0}


\section{Mobility Data Representativeness}

\subsection{Data Description}
In this study, we leverage a large and longitudinal dataset of anonymized individual GPS location mobility data provided by Cuebiq. This is an anonymous, privacy-enhanced, and high-resolution mobile location pings for more than 15 million active users in the United States. All devices within the study opted-in to anonymous data collection for research purposes under the General Data Protection Regulation (GDPR) and the California Consumer Privacy Act (CCPA) compliant framework. This data consists of anonymized user IDs, location coordinates, start time, and dwell time for all users in the United States who have opted to share their location via certain partner applications. In total, there are 2,990,309 data points for 208,344 users whose data are available for the state of Colorado from November 2021 to June 2022. The stop locations were aggregated to a Census block group (CBG) level to preserve privacy. 

\subsection{Robustness to threshold night time for sampling users and home location estimation}
\begin{figure}[h]
    \centering
    \includegraphics[width=0.8\linewidth]{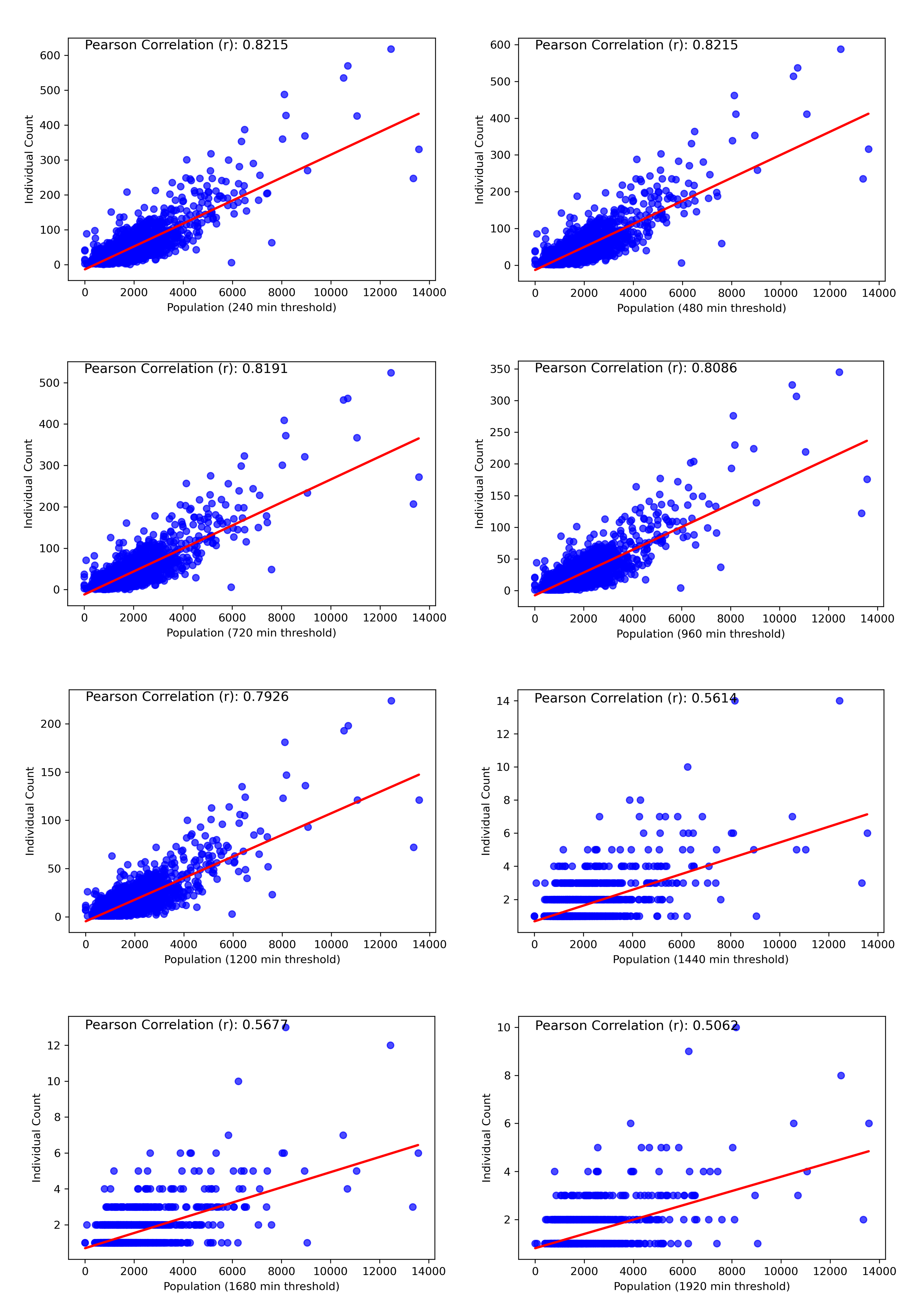}
    \captionsetup{singlelinecheck=off} 
    \caption[Representativeness of the sample with different night time thresholds.]{\textbf{Representativeness of the sample with different night time thresholds.} \\ 
    Pearson correlation between user counts in each census block group for the sampled data at each night time threshold with total population in the CBG remains the same for 240m and 480m ($r =  0.8215$) but begins to drop as thresholds increase, from $r =  0.8215$ to $r =  0.5672$.}
    \label{fig:Figure S1}
\end{figure}

We sampled individuals based on nighttime location data at the Census Block Group (CBG) level, including only those who spent at least the specified threshold duration (240, 480, 720, 960, 1200, 1440, 1680, or 1920 minutes) in one or more CBGs. To ensure accurate home location estimation, we selected only those individuals whose data was available for a minimum of five nights per week during the pre-disaster period (November 2021). The sampling percentage for each threshold was calculated by dividing the number of observed users per CBG by the total population of the corresponding CBG, multiplied by 100. To assess the representativeness of the data at each threshold, we computed the correlation between the total population of each CBG and the number of users sampled from that CBG Figure \ref{fig:Figure S1}. For the selected threshold of 480 minutes, the correlation coefficient was high ($r = 0.8215$), indicating that the dataset is representative of the population in each CBG. The home CBG for each individual was estimated as the CBG where they had the highest cumulative dwell time (in minutes) during the four weeks preceding the disaster (disaster dates: December 30, 2021, to January 1, 2022). Nighttime was defined as the hours between 8:00 PM and 7:00 AM.
   

\subsection{Population and income representativeness}
\begin{figure}[h]
    \centering
    \includegraphics[width=0.8\linewidth]{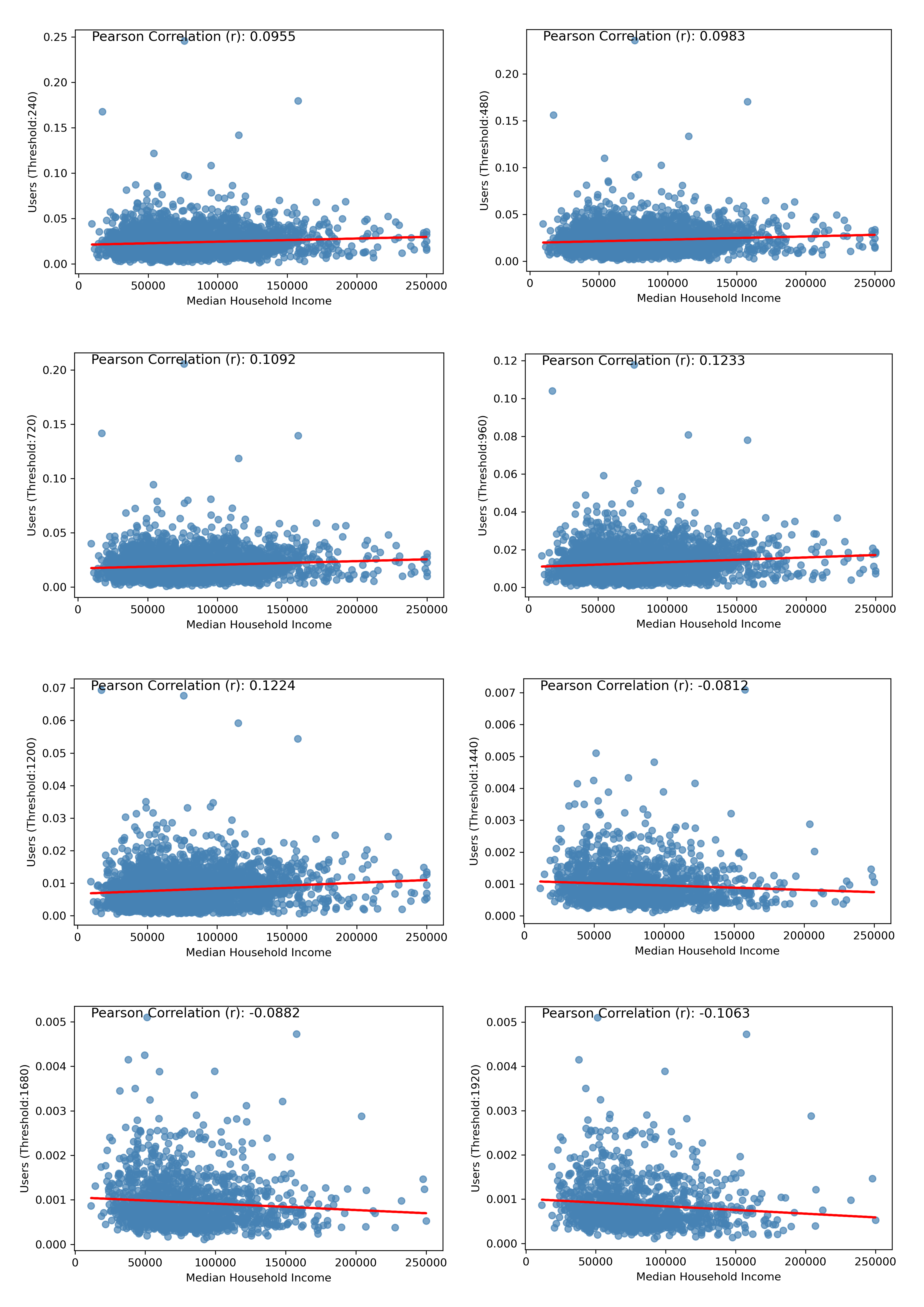}
    \captionsetup{singlelinecheck=off} 
    \caption[Representativeness of income groups in the sampled users.]{\textbf{Representativeness of income groups in the sampled users with different night time thresholds.} \\ 
    Pearson correlation between the users in each census block group in the sampled data and the median household income for the CBG's is close to zero across almost all thresholds. Threshold 480m: $r =  0.0983$.}
    \label{fig:Figure S2}
\end{figure}

\begin{figure}[h]
    \centering
    \includegraphics[width=0.8\linewidth]{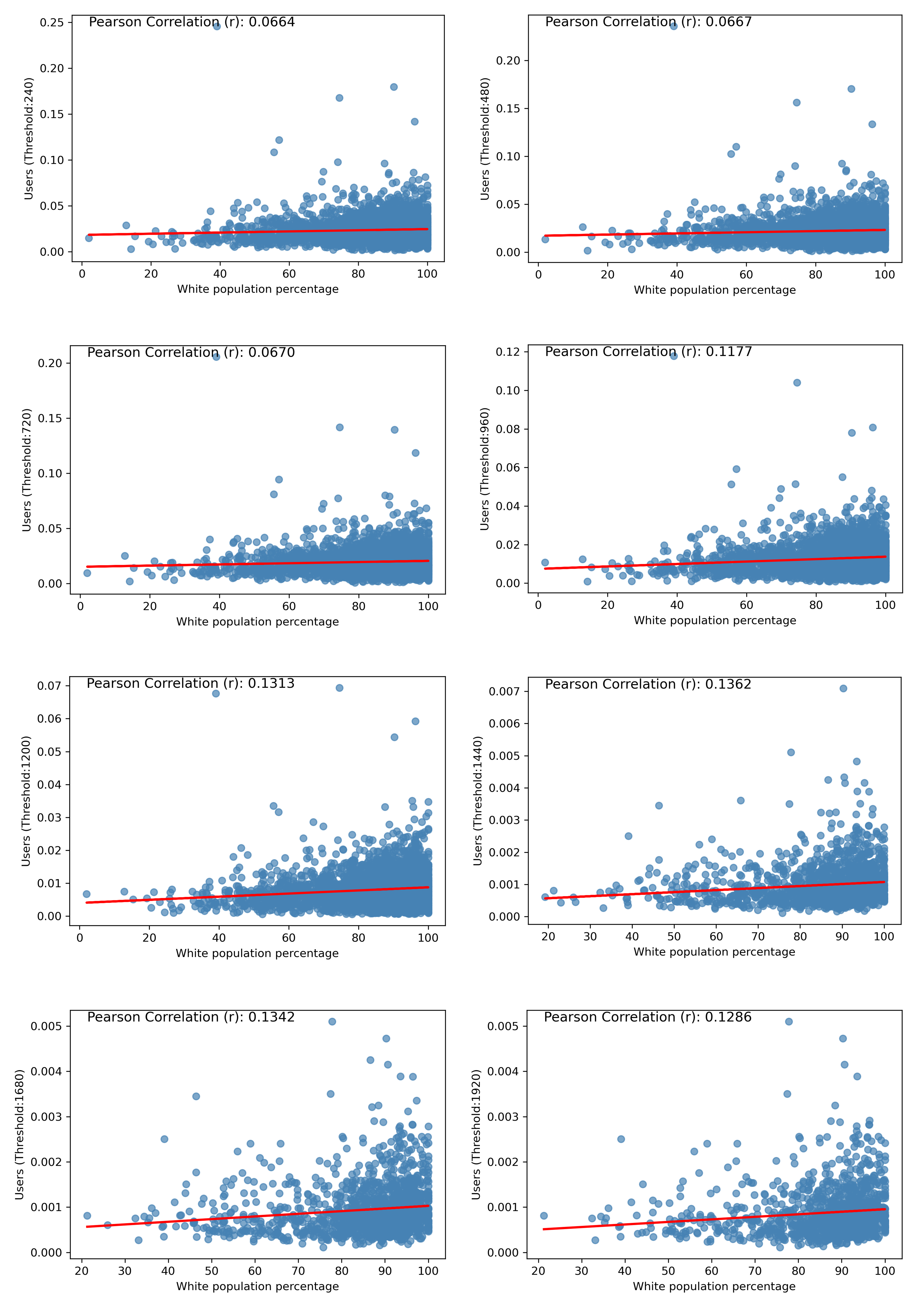}
    \captionsetup{singlelinecheck=off} 
    \caption[Representativeness of racial groups in the sampled users with different night time thresholds.]{\textbf{Representativeness of racial groups in the sampled users with different night time thresholds.} \\ 
    Pearson correlation between the users in each census block group in the sampled data and the median household income for the CBGs is close to zero across almost all thresholds. However, it ranges from $r =  0.0664$ for 240m with a small increase to $r =  0.0667$ for 480m and then keeps increasing to $r =  0.1286$ for 1920m.}
    \label{fig:Figure S3}
\end{figure}

In addition to checking the sampling data, it is also important to check the representativeness of income groups and racial groups in the sampled data. Figure \ref{fig:Figure S2} and Figure \ref{fig:Figure S3} show the correlation of the sample population for each nighttime threshold with the income and race proportions, respectively. Demographic data like population, race percentage, and income percentage were obtained from the American Community Survey (ACS) of 2019. A lower correlation would indicate that the dataset is balanced. The plots for the threshold of 240 - 480 minutes show that the sample rate has a negligible correlation with the income and race proportions, further indicating that the chosen sample of 480 minutes ($R_{income}=0.0983$, $R_{race}=0.0667$) is balanced. In contrast, for thresholds beyond 480 minutes, the correlations with income and racial proportions increase slightly. For income,  $R_{income}$  peaks at 960 minutes ($R_{income}=0.1233$) before declining and becoming slightly negative at the 1920-minute threshold ($R_{income}=-0.1063$). For race,  $R_{race}$  continues to rise marginally, peaking at 1440 minutes ($R_{race}=0.1362$). These trends suggest that higher thresholds eventually reduce the representativeness of the dataset.

\begin{figure}[h]
    \centering
    \includegraphics[width=0.65\linewidth]{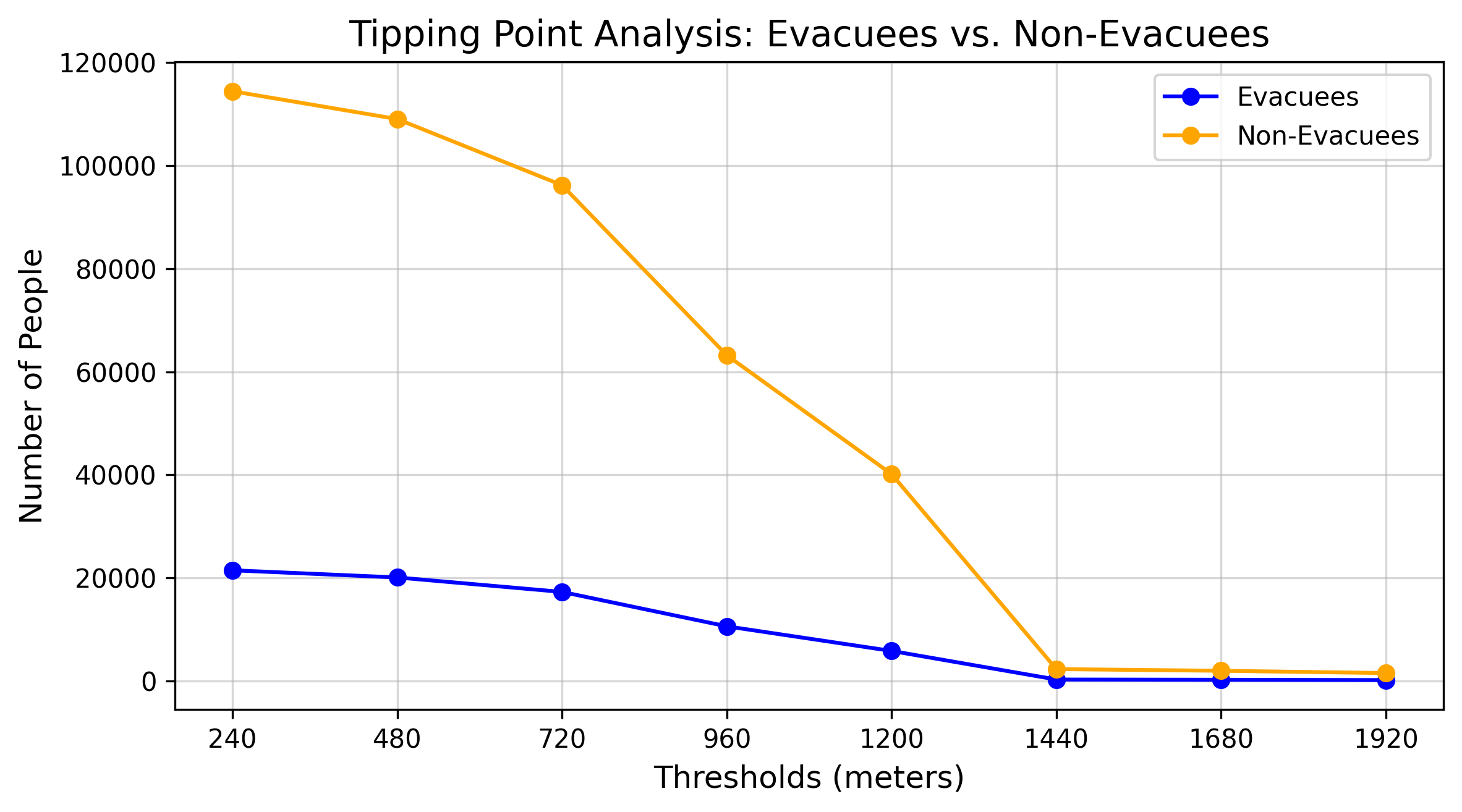}
    \captionsetup{singlelinecheck=off} 
    \caption[Number of evacuees and non evacuees based on various night time thresholds ]{\textbf{Number of evacuees and non evacuees based on various night time thresholds} \\ 
    Evacuees (blue line): The number of evacuees consistently decreases as the threshold increases, with a steep decline after 720 minutes and reaching minimal levels beyond 1440 minutes, Non-Evacuees (orange line): The number of non-evacuees also decreases as the threshold increases, but at a faster and more dramatic rate compared to evacuees, also beyond 720 meters indicating that the use of 480 minute threshold is the best option.}
    \label{fig:Figure S4}
\end{figure}

\section{Evacuation Behavior}
\subsection{Detecting evacuees and non-evacuees}
Employing a similar approach using night-time data, we estimate the weekly night-time stay locations (CBG) for 3 days during the disaster and 22 weeks after the disaster. Each week, the CBG where each individual spent maximum time during the night-time period was labeled as their home location for that week. An individual was labeled as evacuated if they spent the most time (more than 480 minutes) in a CBG other than their home CBG during nighttime in the three days following the disaster. 
It was important to check the robustness of the results on evacuation behavior modeling against the selection of the minimum time threshold. The graph in Figure \ref{fig:Figure S4} demonstrates a considerable drop in evacuee counts beyond 720 minutes duration, indicating the suitability for 480 minutes for capturing nighttime location patterns. 
To enable a more robust and longitudinal analysis, we include individuals whose data is observed continuously and is not missing for at most three consecutive weeks. To focus our analysis on the impacts of the fire, we focus on individuals whose home location was within a 100 kilometer radius of the epicenter of the fire. 
Our sampled data used for further analysis consists of 122,432 users of which 13,970 were labeled as evacuees. 

\subsection{Social determinants of evacuation}
To analyze the social and spatial determinants of evacuation dynamics we compute the evacuation rate for each census block group. Evacuation rate can be defined as the proportion of individuals in the data who evacuated in the first 3 days after the fire, from their home census block groups (CBGs). 
The logistic regression model employed to understand the factors associated with evacuation decisions with the specification: $Pr(y_u = 1) \sim logit^{-1}[G_u+SD_u+B_u]$, where $y_u=1$ indicates the user $u$ evacuated and $y_u=0$ otherwise. Independent variables include $G_u$, $SD_u$, and $B_u$, which are groups of geographical, sociodemographic, and behavioral characteristics, respectively. To ensure the robustness of the results it was important to test the model for data sampled for various night time thresholds. Table~\ref{tab: Logistic_evac_480} to \ref{tab: Logistic_evac_1440} show the beta coefficients, significance, and confidence intervals for the logistic regression of evacuation decision for thresholds of 480m, 720m, 960m, 1200m, and 1440m. 
The geographical characteristics in the independent variables of the regression model include distance from fire and distance between the home location and evacuation destination which has been calculated using the distance between the centroid geometry of the home and destination CBG polygon. Sociodemographic characteristics variables selected for this analysis include population density, unemployment rate, median income, total housing units, and the percentages of White, Black, and Asian populations, as well as the percentage of the educated population. These data were sourced from the American Community Survey (ACS). Behavioral characteristics include an individual's pre disaster mobility behavior: entropy $H$ and radius of gyration $R_g$. 

\subsection{Pre-disaster Mobility Behavior}
To assess the pre-disaster mobility behavior of selected evacuees, we compute two key metrics, Entropy $H$ and Radius of Gyration $(r_g)$ using recorded mobility data from November 1st to 30th, 2021, prior to the disaster. For this calculation, we use all stop points of the sampled individuals for night time thresholds of (480m, 720m, 960m, 1200m, 1440m). These metrics capture the spatial and temporal patterns of individual movement across visited locations, defined at the level of census block groups (CBGs). 

For each individual, we calculate the number of unique locations visited, based on distinct timestamps representing separate visits, as well as the total dwell time in minutes at each location. To achieve this we use the pandas' scipy.stats.entropy  function. Here, entropy $H$ quantifies the diversity of an individual’s movement by assessing how evenly distributed their total dwell time is across all visited locations. Higher entropy values indicate greater movement diversity and lower predictability, while lower values reflect more concentrated patterns and higher predictability. Mathematically, Shannon entropy is computed as $$H = -\sum (p_k \log(p_k))$$, where $p_k$ represents the proportion of total dwell time at a specific location (in our case all unique locations) relative to the overall dwell time across all locations. This provides a robust metric for evaluating the variability in an individual’s location choices over time.

The radius of gyration is defined as a measure of the characteristic distance traveled by an individual on a regular basis. It captures the spread of an individual’s trajectory around their center of mass. ($r_g = \sqrt{\frac{1}{N} \sum_{i=1}^{N} \left| \mathbf{r}_i - \mathbf{r}_{\text{cm}} \right|^2}$).

In this analysis, we use a weighted version of the standard radius of gyration $r_g$: $$r_g = \sqrt{\frac{1}{N} \sum_{i \in L} n_i (r_i - r_{cm})^2}$$, where L represents the set of locations, $r_i$ the geographic coordinates of location i, $n_i$ the time spent at location i, and $r_{cm}$ the center of mass of all visited locations. The formula used in this analysis differs from the standard radius of gyration formula  because it includes  $n_i$ , which represents weights associated with the positions $i \in L$ which represents the frequently visited locations. 

Together, these metrics provide a detailed understanding of evacuees’ mobility patterns, capturing both the diversity and spatial extent of their movements before the disaster. 

\begin{figure}[ht!]
\centering
\subfloat[\textbf{Relationship between individuals' radius of gyration and income of their home CBG.}]{
    \includegraphics[width=.45\linewidth]{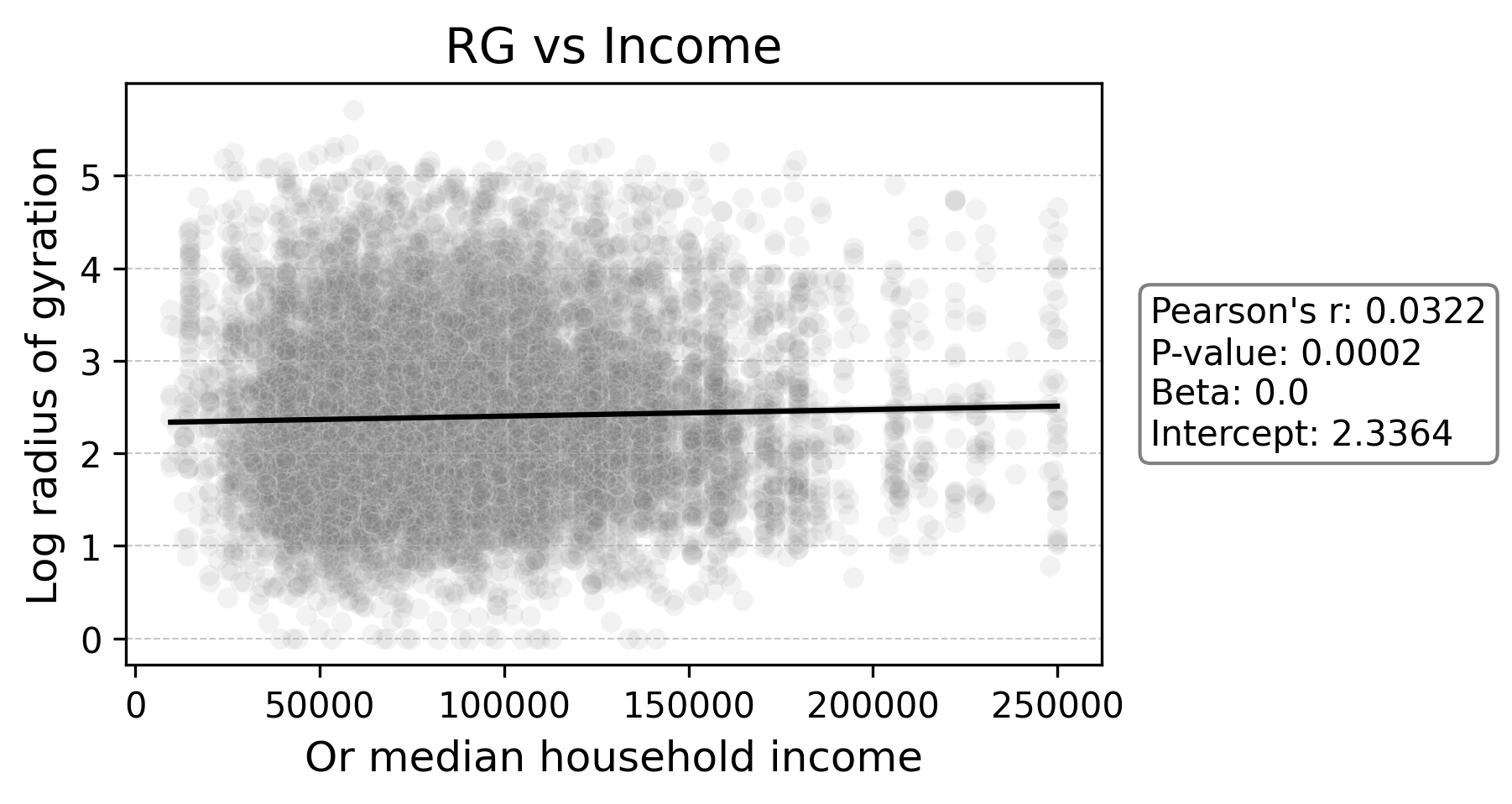}} \hspace{0.05\linewidth}
\subfloat[\textbf{Relationship between individuals' radius of gyration and white population in their home CBG.}]{
    \includegraphics[width=.45\linewidth]{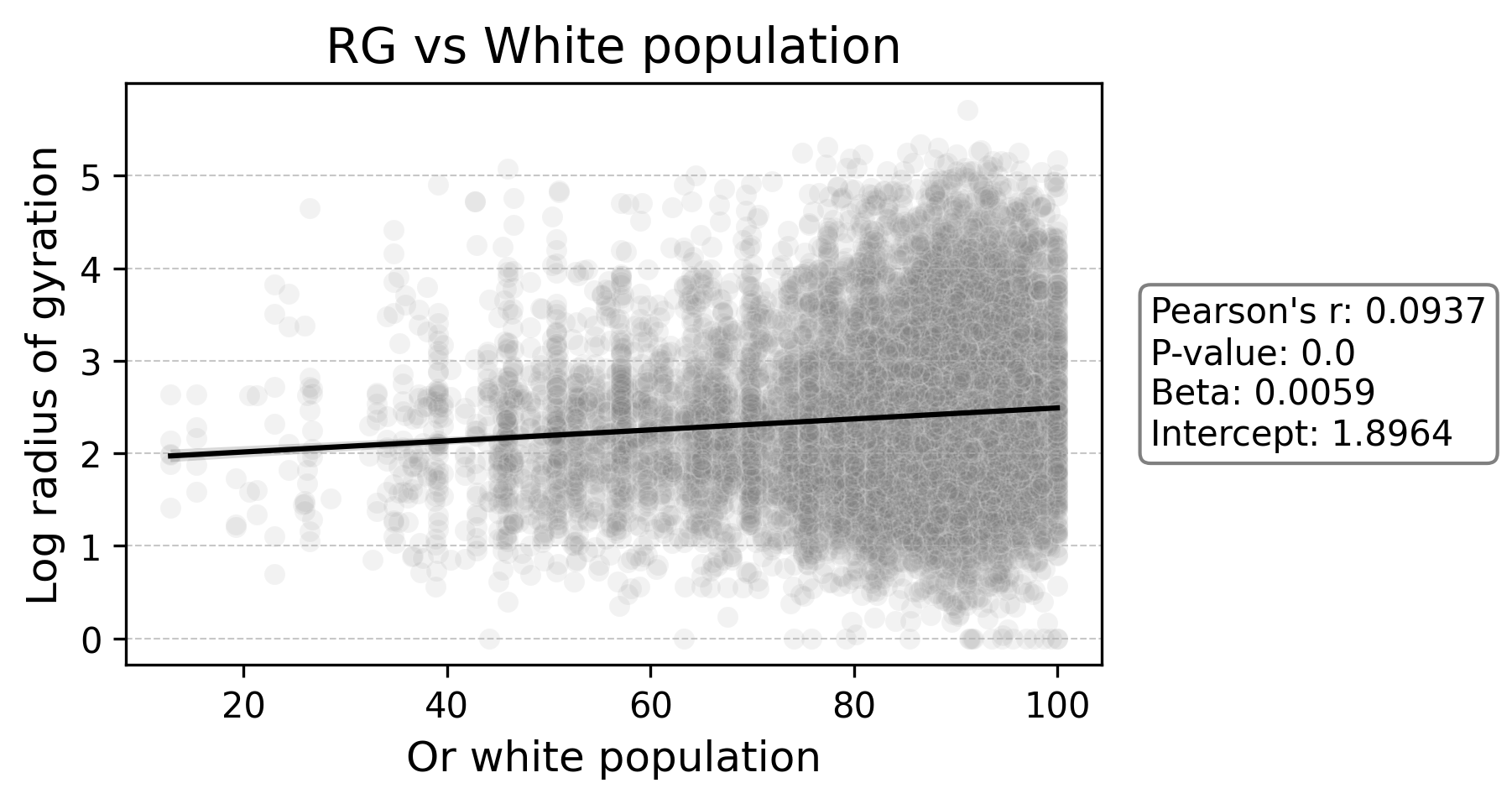}} \\[1em]

\subfloat[\textbf{Relationship between individuals' entropy and income of their home CBG.}]{
    \includegraphics[width=.45\linewidth]{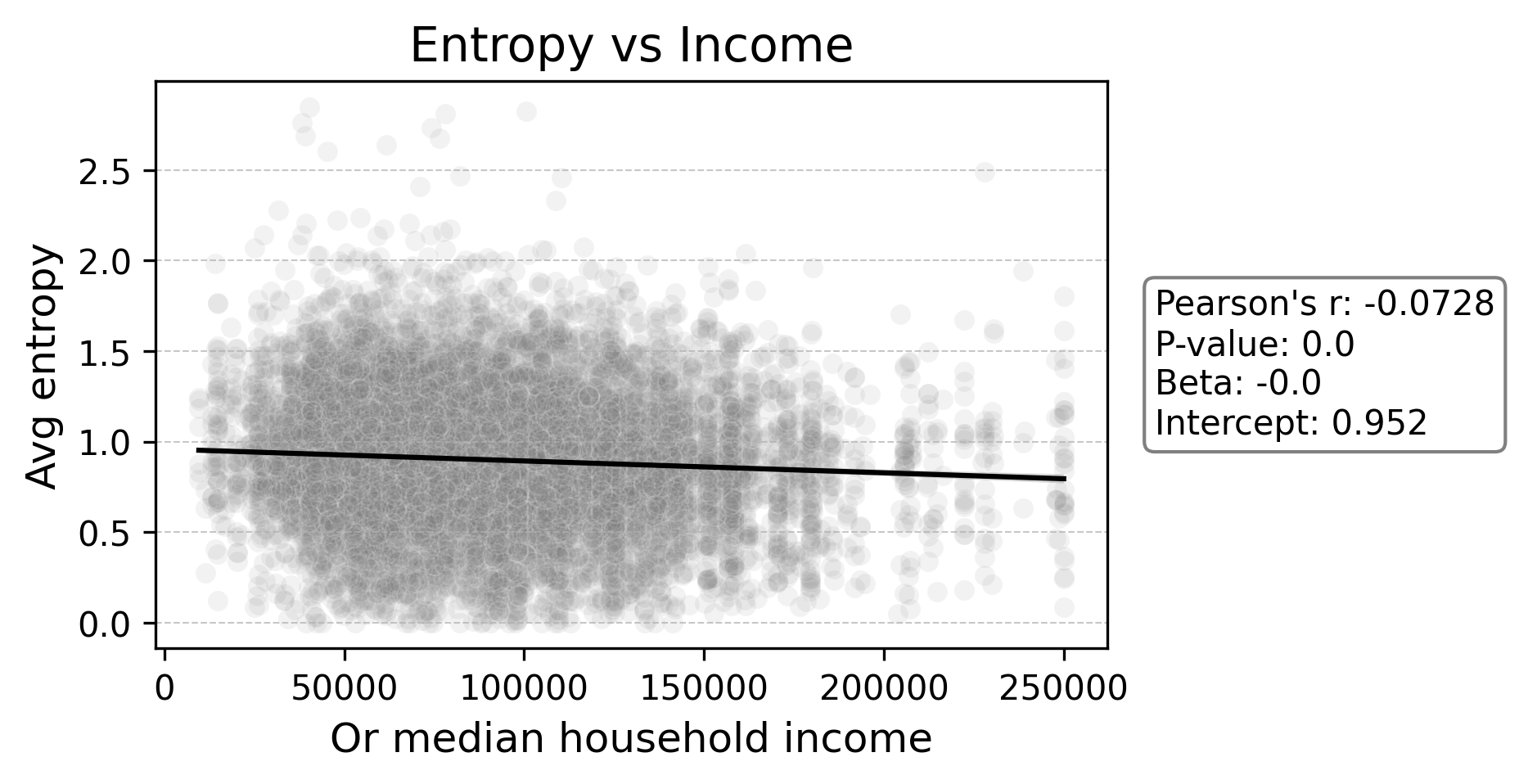}} \hspace{0.05\linewidth}
\subfloat[\textbf{Relationship between individuals' entropy and white population in their home CBG.}]{
    \includegraphics[width=.45\linewidth]{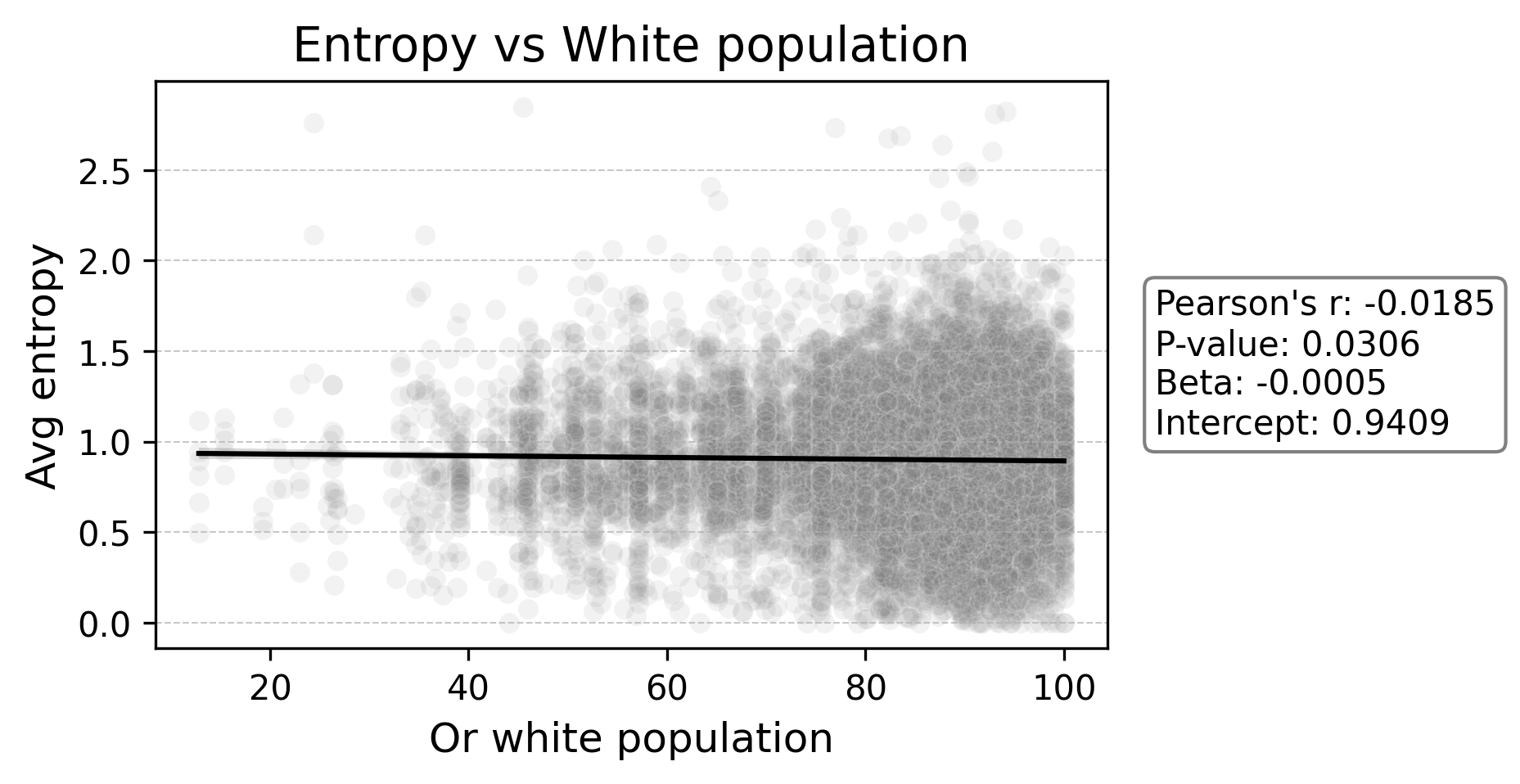}} \\[1em]

\captionsetup{singlelinecheck=off} 
\caption[Relationship between individuals' predisaster mobility variables and their demographic characteristics.]{\textbf{Relationship between individuals' predisaster mobility variables and their demographic characteristics.} \\ 
(a) Pearson correlation between an individual's radius of gyration and median income of their home CBG is significant and close to zero ($r = 0.032$). 
(b) Pearson correlation between an individual's radius of gyration and white population of their home CBG is significant and close to zero ($r = 0.093$). 
(c) Pearson correlation between an individual's entropy and median income of their home CBG is significant and close to zero ($r = -0.072$). 
(d) Pearson correlation between an individual's entropy and white population of their home CBG is significant and close to zero ($r = 0.030$). }
\label{fig:Figure S5}
\end{figure}

\begin{figure}[ht!]
\centering

\subfloat[\textbf{Correlation: radius of gyration demographic characteristics}]{
    \includegraphics[width=.6\linewidth]{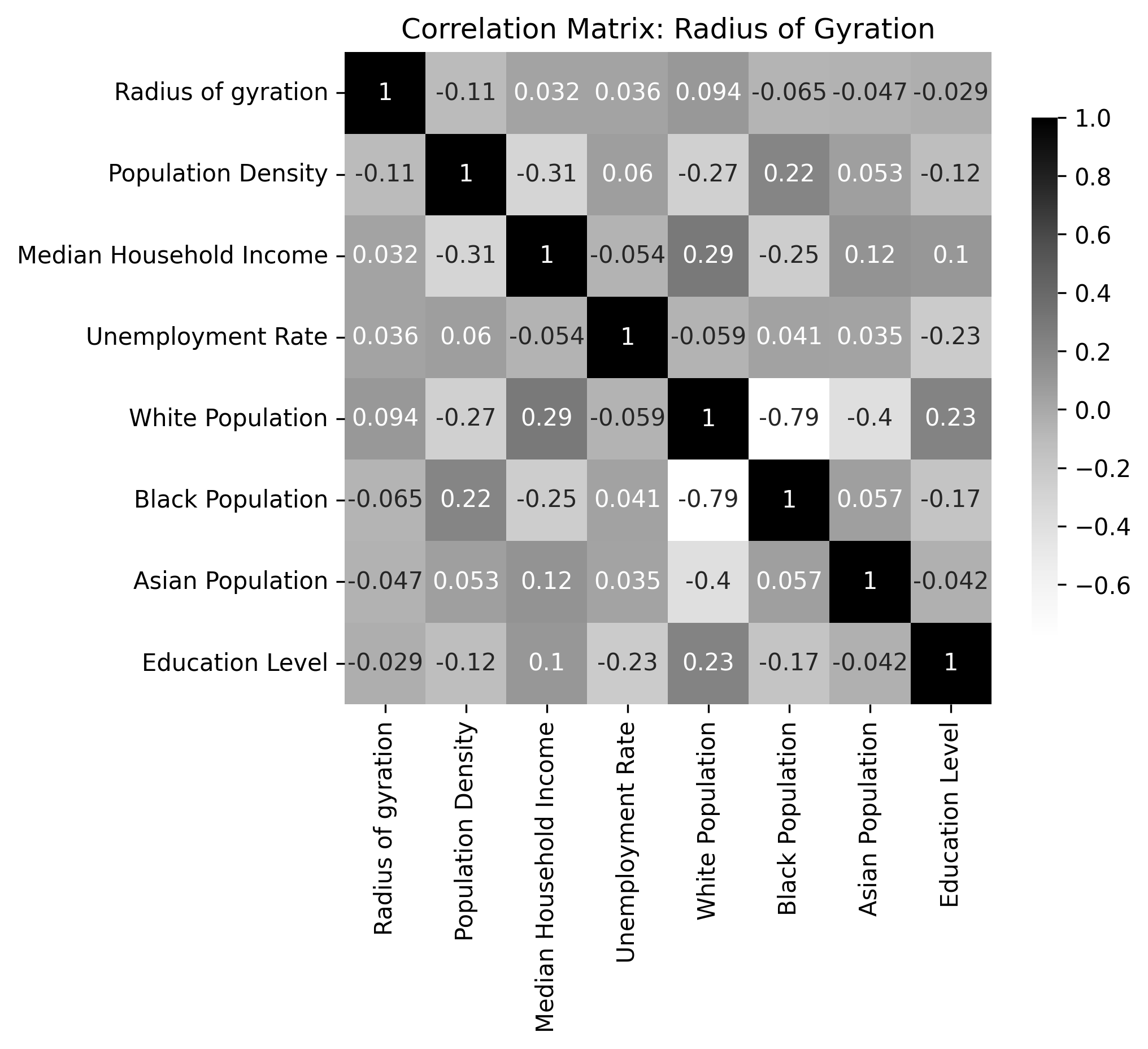}} \\[1em]

\subfloat[\textbf{Correlation: entropy and demographic characteristics}]{
    \includegraphics[width=.6\linewidth]{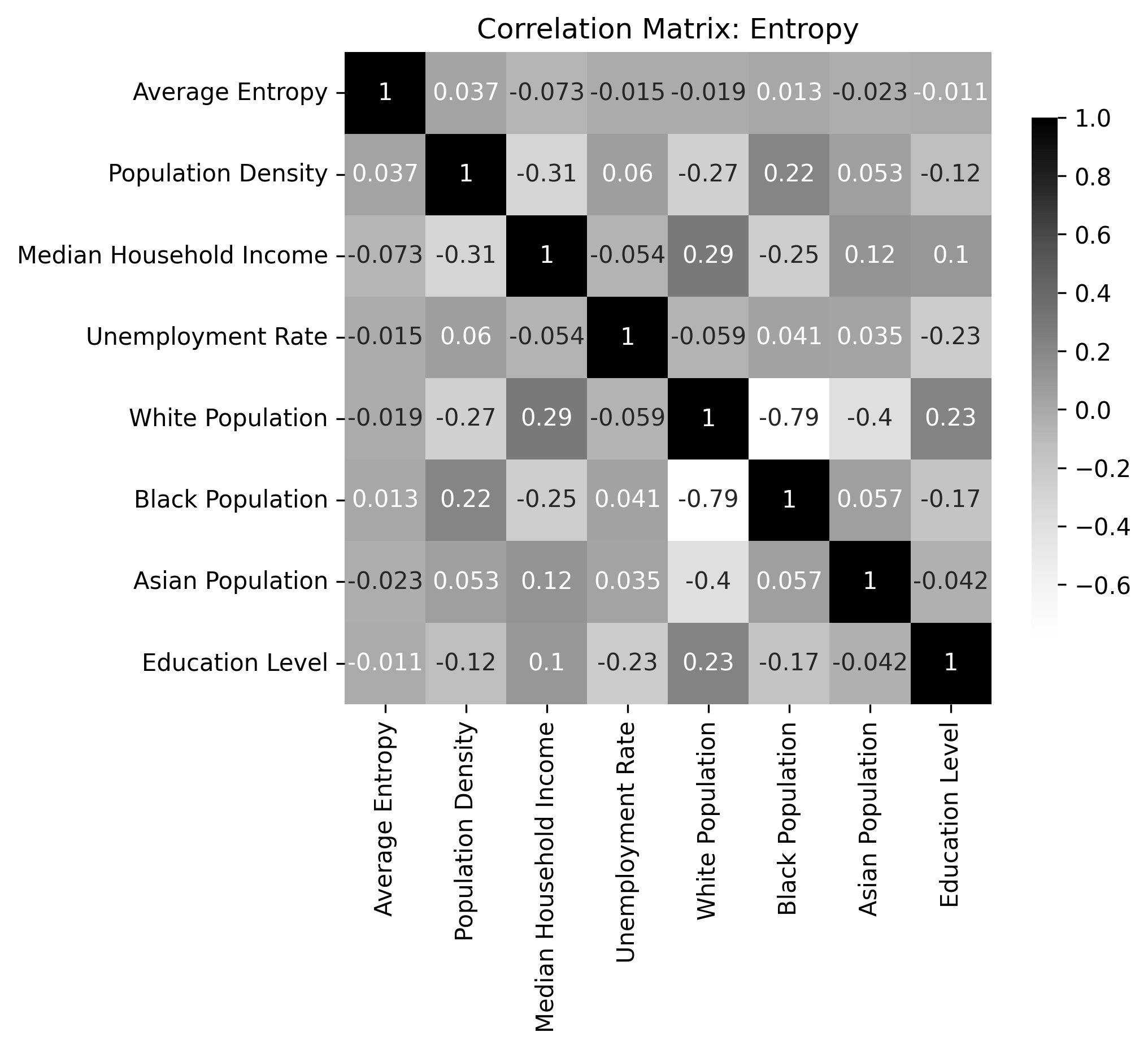}}

\caption[Correlation between predisaster mobility variables and demographic characteristics.]{
    \textbf{Correlation between individuals' predisaster mobility variables and their demographic characteristics.}  
    Most correlations between individuals’ radius of gyration or entropy and various demographic and socioeconomic variables are weak, suggesting minimal linear relationships between predisaster mobility variables and these characteristics.}
\label{fig:Figure S6}
\end{figure}

It was important to gain a deeper understanding of the demographic characteristics associated with variations in pre-disaster mobility behavior (measured by entropy and radius of gyration), we analyze the correlation between race and income with both entropy and radius of gyration for each individual (\ref{fig:Figure S5}). Additionally, we construct a correlation matrix that includes all demographic variables alongside pre-disaster mobility behavior metrics to identify broader patterns and relationships, illustrated in \ref{fig:Figure S6}. The correlation coefficients for all demographic variables are close to zero indicating that there is no significant relationship between pre-disaster mobility behavior metrics (Entropy and Radius of Gyration) and demographic characteristics.

\section{Modeling social homophily between origin and evacuation destinations.}
\subsection{Social homophily measures}
\begin{figure}[h]
    \centering
    \includegraphics[width=0.75\linewidth]{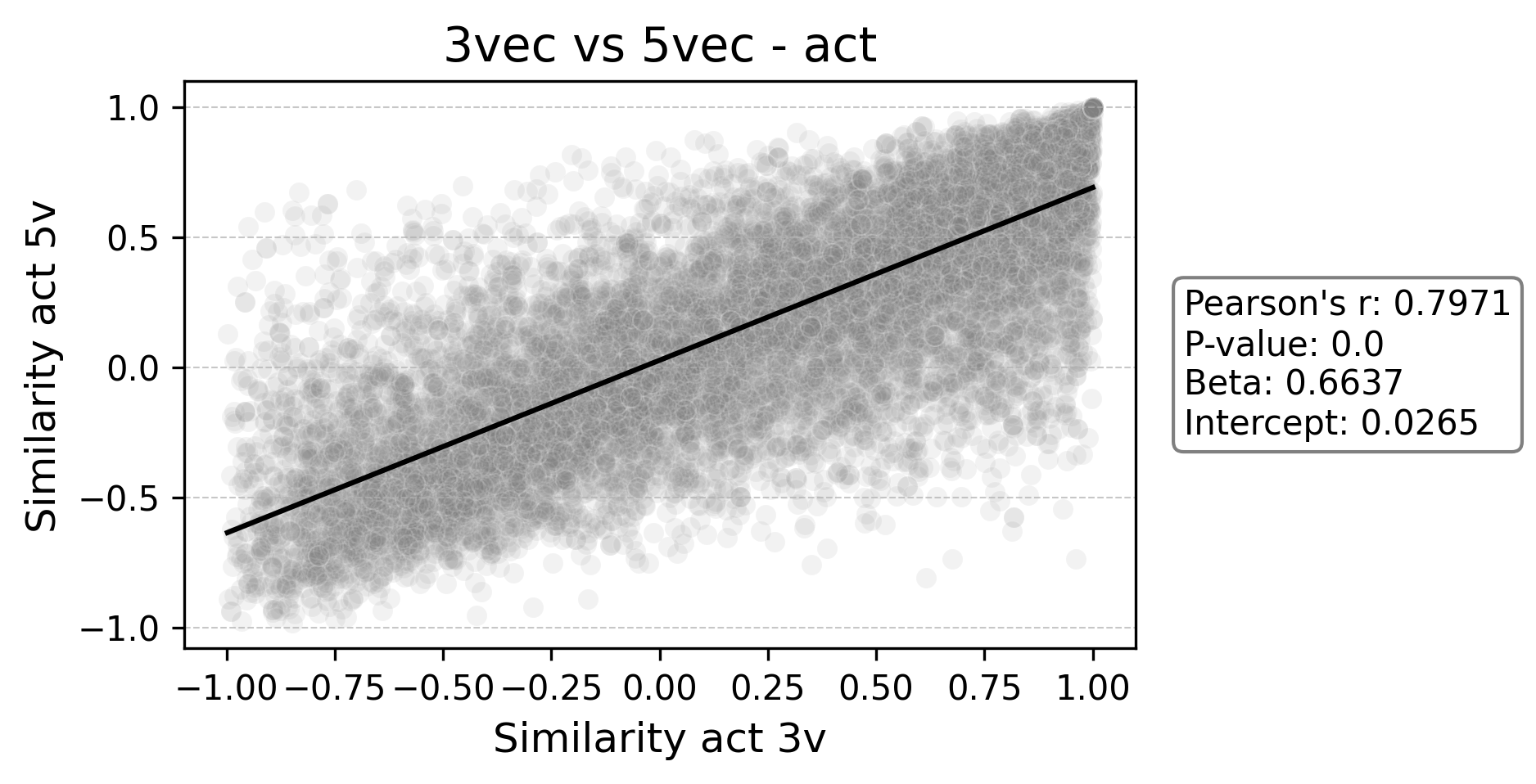}
    \captionsetup{singlelinecheck=off} 
    \caption[Variation in similarity value for actual destinations using 5-feature versus 3-feature vectors.]{\textbf{Variation in similarity value for actual destinations using 5-feature versus 3-feature vectors.} \\ 
    Pearson correlation between the similarity values computed using 5-feature vectors and those computed using 3-feature vectors is very high ($R = 0.7971$), indicating a strong alignment between both approaches.}
    \label{fig:Figure S7}
\end{figure}

We compute the social homophily between an individual's home and evacuation destination using two measures namely Similarity and Connectedness. The similarity is the sociodemographic similarity between home and evacuation destination CBG's calculated using cosine similarity. Cosine similarity is the cosine of the angle $\theta$ between two vectors v and w in a n-dimensional space and it ranges from -1 to 1. Larger the angle between the vectors, lower will be similarity. The closer the cosine value to 1, the more similar the two vectors are. $$Similarity_{vw} = cos(\theta) = \frac{v . w}{||v||||w||}$$. 5 sociodemographic vectors were selected for computing similarity and these include population density, median household income, race proportions in the population, unemployment percentage and education percentage. However, it was also important to ensure that the variation in similarity value with different set of vectors is comparable to the value computed before. Therefor similarity was also computed for each home and evacuation destination using three vectors which include population density, median household income and race variables. \ref{fig:Figure S7} shows the correlation plot between the similarity values of actual destinations computed with three versus five vectors which indicates a high correlation(($R = 0.7971$), and a strong alignment between both approaches. 

\subsection{Simulating Destinations for exposed homophily}
Exposed homophily, in this study, refers to the extent to which individuals’ choices, such as evacuation destinations are shaped by external physical factors (e.g., geographic distance, population size) rather than social and behavioral preferences such as friendship connections or socioeconomic characteristics. 

To quantify exposed homophily and identify destinations that individuals would have selected only based on their exposed homophily, we constructed a null model to simulate evacuation destinations for each individual. For each individual, we generated a set of origin-destination pairs by employing a weighted random sampling method, ensuring that all simulated destinations fall within the same distance decile bin as the individual’s actual destination therefore controlling on evacuation distance ($d_{ij}$). This also approximates the expected destination choice if the decision completely depended on the gravity model, which posits that destinations with larger populations ($M_j$) exert a stronger pull. To achieve this we assign weights to each origin-destination pair proportional to the population of the destination, defined as $p_{ij} = M_j/d_{ij}^2$. Entries with zero probabilities were excluded to ensure valid sampling, and one destination per individual was sampled probabilistically. 

\subsection{Comparing actual homophily with exposed homophily}
\begin{figure}[h]
    \centering
    \includegraphics[width=0.75\linewidth]{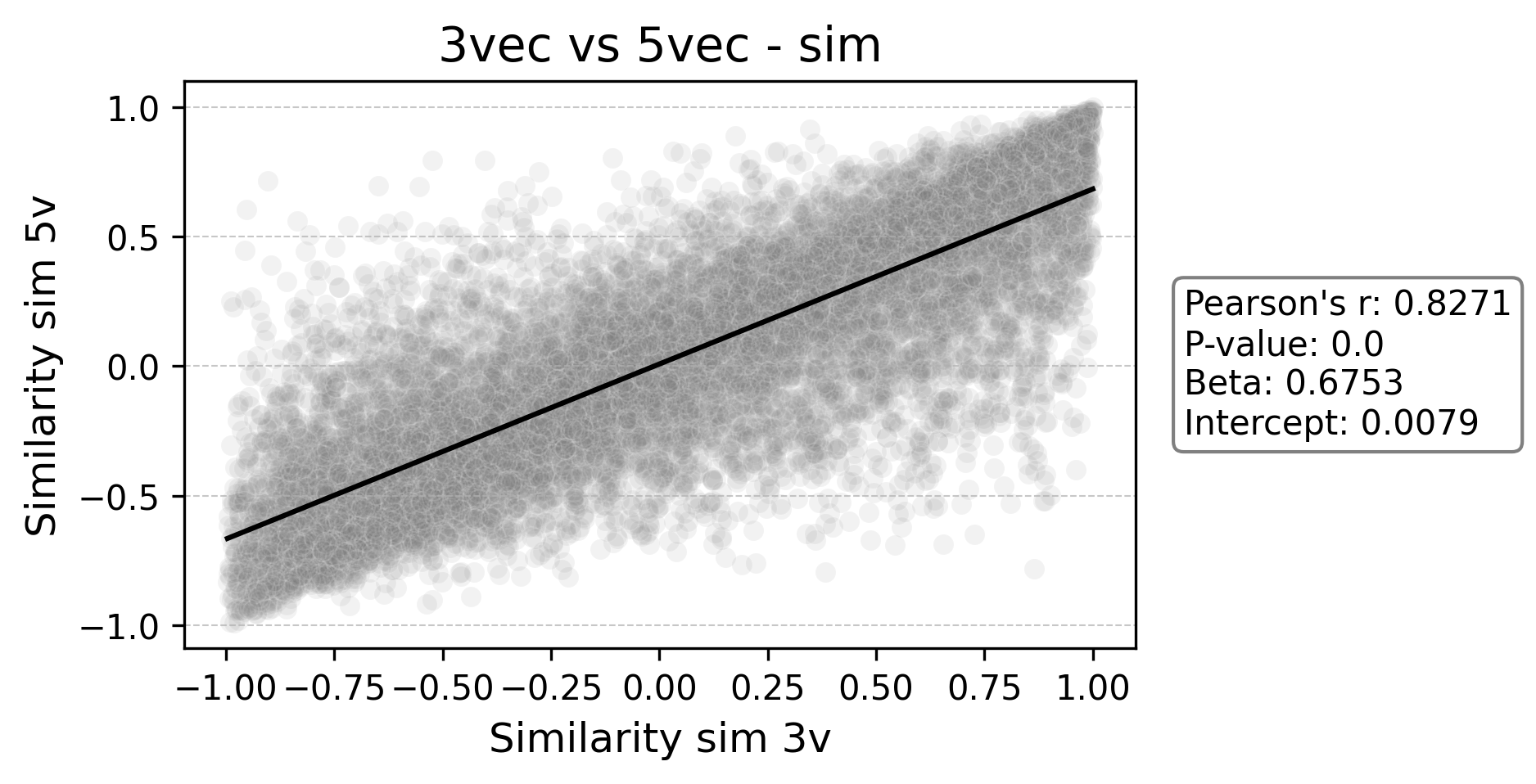}
    \captionsetup{singlelinecheck=off} 
    \caption[Variation in similarity value for simulated destinations using 5-feature versus 3-feature feature vectors.]{\textbf{Variation in similarity value for simulated destinations using 5-feature versus 3-feature vectors.}\\ Pearson correlation between the similarity computed using 5-feature vectors and 3-feature vectors similarity values is very high with ($R = 0.8271$) also indicates strong alignment between both approaches.
    }
    \label{fig:Figure S8}
\end{figure}

To obtain a better understanding of the computation method of social similarity using cosine similarity we re-computed similarity for both actual and simulated destinations using 5 vectors namely population density, median income, race, education, and income; and 3 vectors namely population density, median income, and race. \ref{fig:Figure S8} shows the Person's correlation between values of similarity computed with 3 vectors on x-axis and 5 vectors on y-axis simulated destinations ($r = 0.8271$), indicating strong alignment between the two approaches similar to the case of actual destinations.

\begin{figure}[h]
    \centering
    \subfloat[\textbf{Connectedness $C_{ij}$ for actual (orange box plots) and simulated (grey box plots) destinations.}]{
        \includegraphics[width=0.45\linewidth]{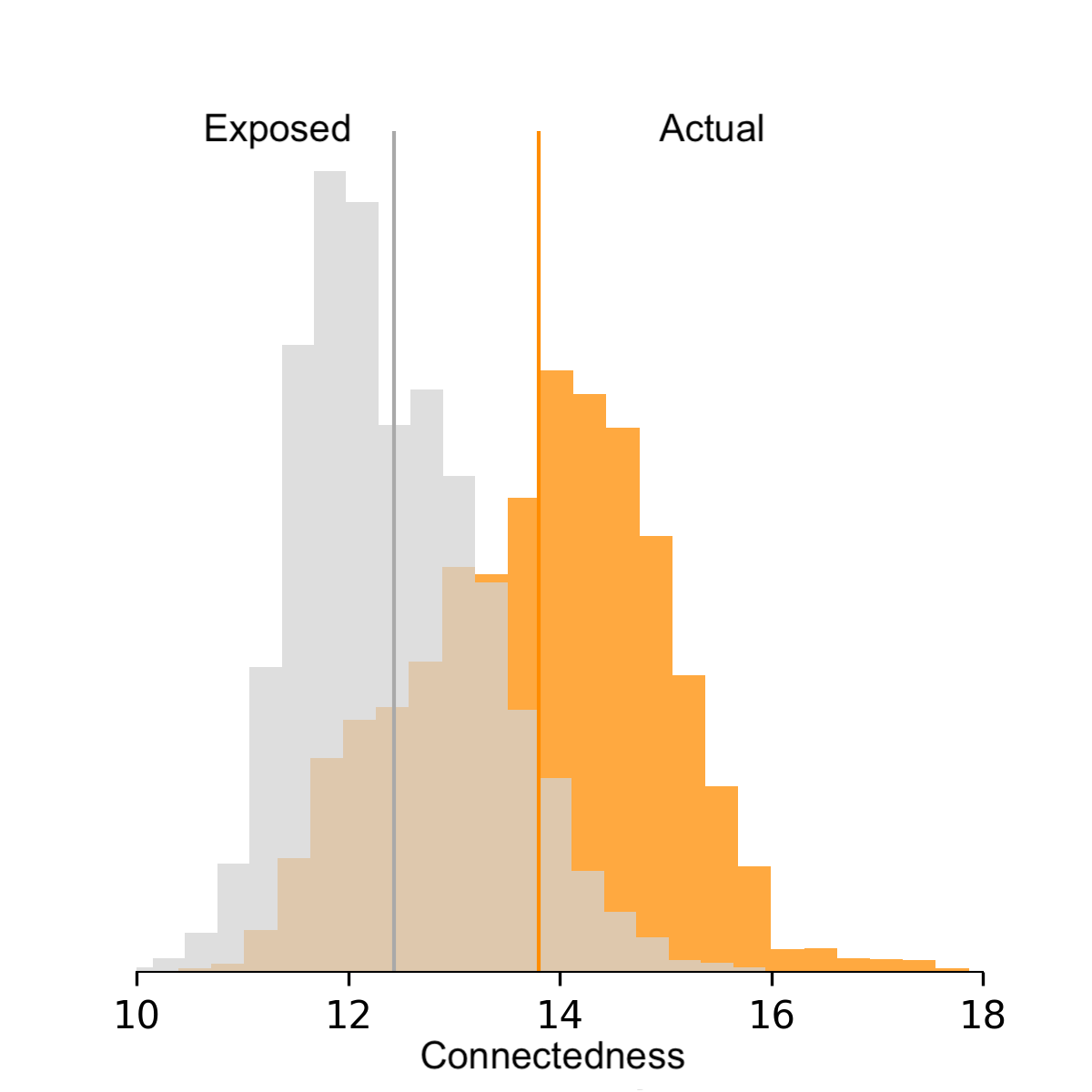}
    } \hspace{0.05\linewidth}
    \subfloat[\textbf{Similarity $S_{ij}$ measured using 5 vectors for actual (blue box plots) and simulated (grey box plots) destinations.}]{
        \includegraphics[width=0.45\linewidth]{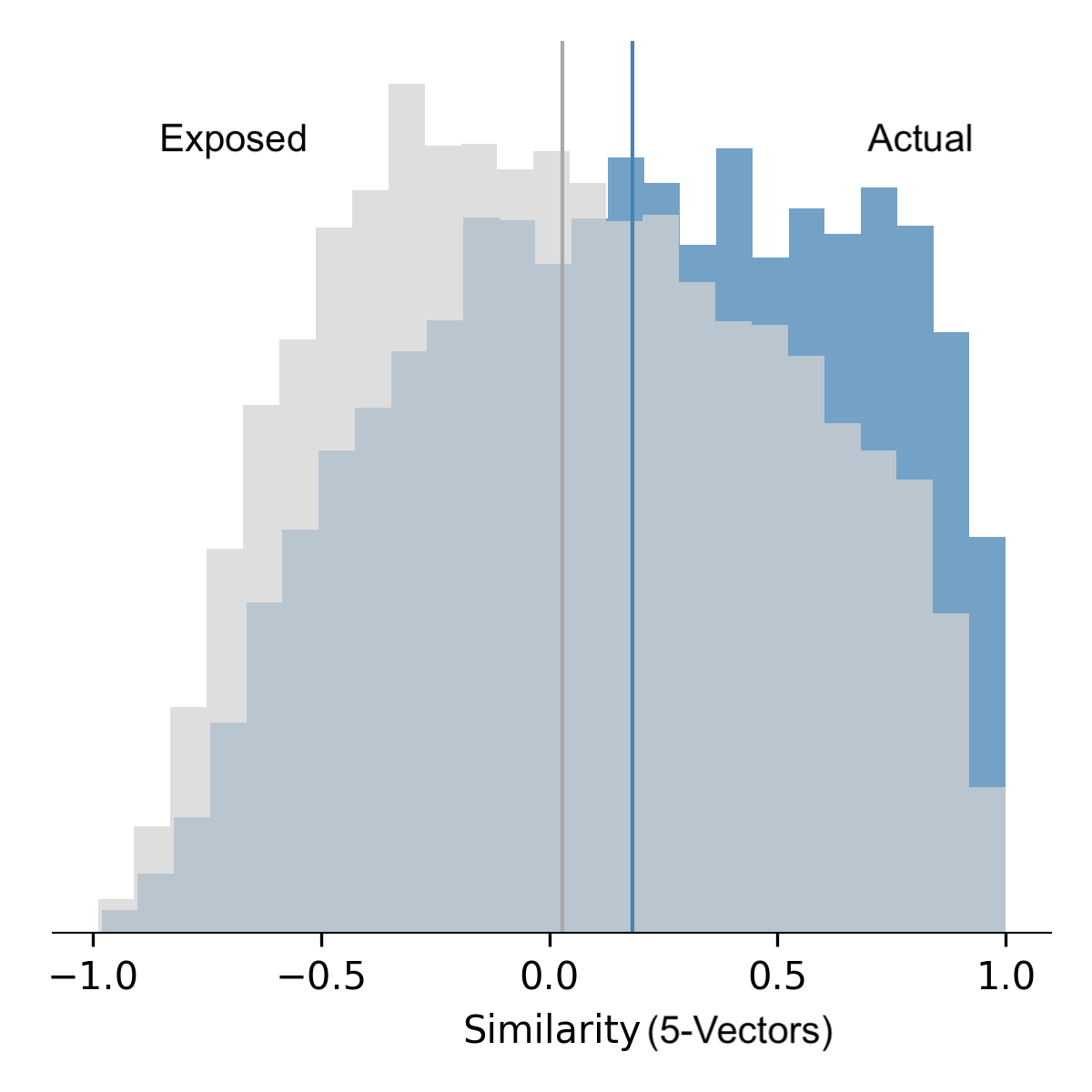}
    }
    \captionsetup{singlelinecheck=off} 
    \caption[Comparison of Connectedness and Similarity for Actual and Simulated Destinations.]{\textbf{Comparison of Connectedness and Similarity for Actual and Simulated Destinations.} \\ 
    (a) Connectedness $C_{ij}$ calculated for both actual and simulated destinations shows higher values for actual destinations compared to the null model ($p \geq 0.05$). 
    (b) Similarity $S_{ij}$ calculated using 5-feature vectors for actual and simulated destinations also shows higher values for actual destinations compared to the null model ($p \geq 0.05$).}
    \label{fig:Figure S9}
\end{figure}

\begin{figure}[h]
    \centering
    \includegraphics[width=0.45\linewidth]{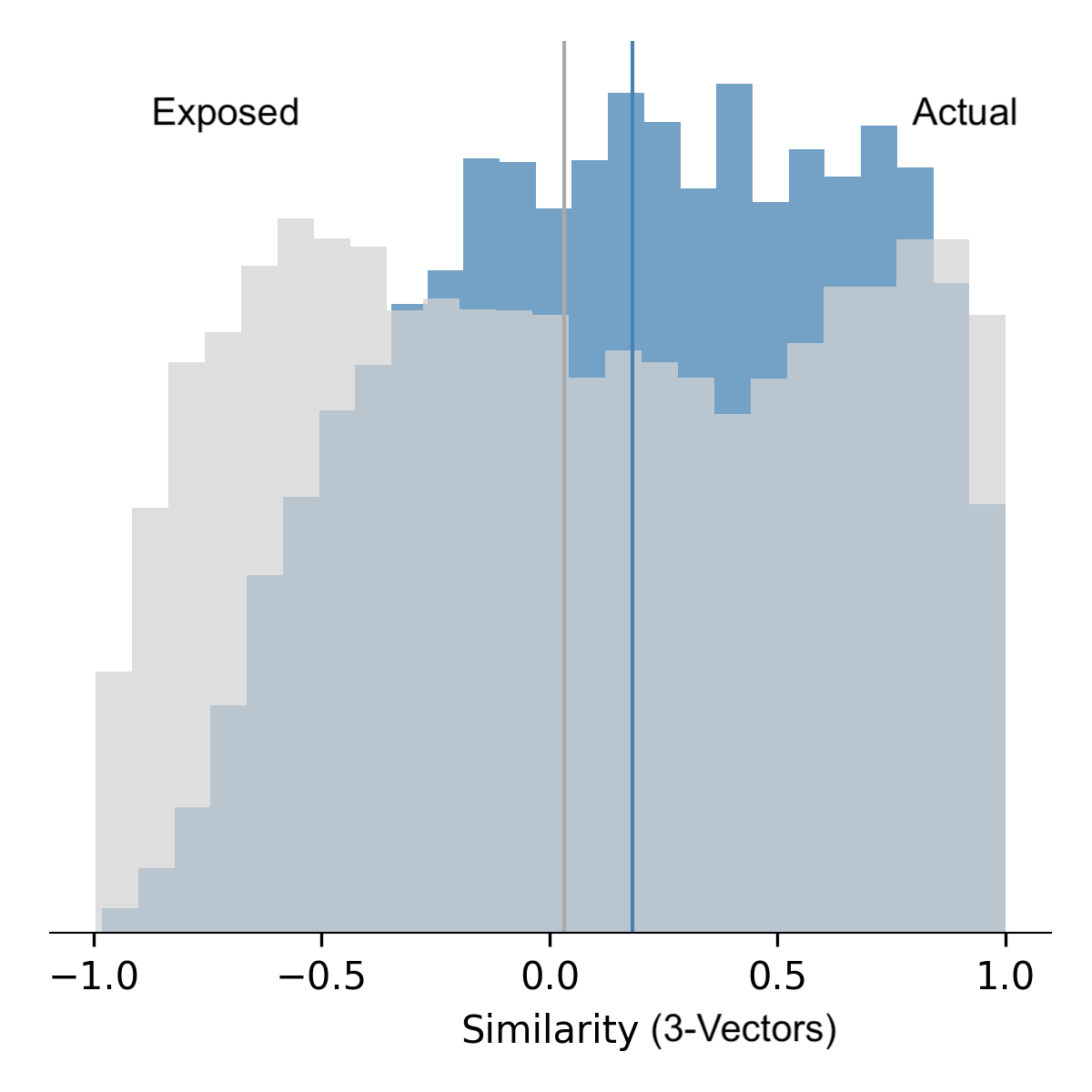}
    \captionsetup{singlelinecheck=off} 
    \caption[Similarity measured using 3-feature vectors $S_{ij}$ for actual and simulated destinations.]{\textbf{Similarity measured using 3-feature vectors $S_{ij}$ for actual and simulated destinations.} \\ 
    Similarity calculated for both actual (blue box plots) and simulated (grey box plots) destinations shows higher connectedness for actual destinations compared to the null model ($p \geq 0.05$).}
    \label{fig:Figure S10}
\end{figure}

Finally, we compared the actual and exposed destinations by calculating the mean similarity and connectedness for both, allowing us to assess the extent to which observed destination choices reflected intrinsic homophily versus opportunities to choose a destination based on one's exposed homophily. \ref{fig:Figure S9} represents the histogram for connectedness and similarity between actual and exposed destinations (simulated using 5 vectors) indicating a higher connectedness and similarity in actual destinations as opposed to exposed destinations. It was important to ensure that the results similarity computed using 3 vectors also aligned with above results. Histogram of similarity  (computed using 3 vectors) for exposed and actual destination \ref{fig:Figure S10} aligns with our previous results, also indicating that actual destinations have higher similarity.  

\subsection{Modeling relationship between demographics and social homophily}
Who evacuates to socially homophylic destinations ($S_{ij}, C_{ij}$)? To investigate this, we employed three linear regression models with social similarity and connectedness as dependent variables and sociodemographic variables ($SD_i$)—including White percent, Black percent, and Median income of the origin census block group—as independent variables, controlling for evacuation distance ($d_{ij}$): $S_{ij}, C_{ij} \sim d_{ij} + SD_i$. The results for these models are represented in Tables \ref{tab: Table S9} to \ref{tab: Table S14}, the $\beta$-coefficients for Black population percentage are significant and negative for both connectedness and similarity, whereas the $\beta$-coefficients for White population percentage are significant only for connectedness and positive for both metrics. Median income, although significant in both models, has a negative coefficient for connectedness and a positive coefficient for similarity.

Next, we used three linear regression models to examine differences in similarity and connectedness between actual and simulated destinations ($\Delta S_{ij}, \Delta C_{ij}$): $\Delta S_{ij}, \Delta C_{ij} \sim d_{ij} + SD_i$. Results are represented in Tables \ref{tab: Table S15} to \ref{tab: Table S20}, indicating that the $\beta$-coefficients for  White population percentage are high and significant ($p \leq 0.05$) for both models. The $\beta$-coefficients for Black population percentage, while also significant, are negative in both cases. Median income again exhibits opposing effects, with $\beta_{\text{median income}} = 0.01$ for similarity and $\beta_{\text{median income}} = -0.05$ for connectedness.

\section{Modeling return and displacement patterns.}
In this section, we show how we address the question - Who gets displaced?  and analyze the relationship between social homophily during evacuation and long-term displacement outcomes. For this purpose weekly return rates were tracked for precautionary and directly impacted evacuees using the weekly home location data computed for each individual for 22 weeks after the fire. Return status was coded as binary: 1 for evacuees who returned to their pre-disaster home locations and 0 for those who remained displaced, with percentages calculated weekly over 22 weeks post-disaster. To compute confidence interval we used bootstrapping method for each week's return percentage of precautionary and directly impacted evacuees to show the statistical significance of the difference in return percentages. Then we employed logistic regression model on who $returned = 1$ and was $displaced = 0$ during after 1st week after the fire and after the 20th week of the fire. Independent variables here include the value of connectedness and similarity between each individual's origin and destination on the given week. \ref{tab: Table S21} to \ref{tab: Table S28} show the regression results for these models.


\begin{table}[!htbp] \centering
  \caption{Regression Results for Evacuation binary variable, Threshold: 480}
\begin{tabular}{@{\extracolsep{5pt}}lc}
\hline \hline
& \multicolumn{1}{c}{\textit{Dependent variable: updated\_evacuation\_status}} \\
\cline{2-2}
\\[-1.8ex] & \textbf{Model 1} \\
\hline \\[-1.8ex]
Const & -1.9433 ($p=0.0$) \\
& [-1.9654, -1.9211] \\
Distance From Fire & -0.0519 ($p=0.0$) \\
& [-0.0715, -0.0324] \\
Population Density & -0.089 ($p=0.0$) \\
& [-0.1106, -0.0675] \\
Asian Percent & 0.1021 ($p=0.0$) \\
& [0.0781, 0.1262] \\
Black Percent & -0.0873 ($p=0.0$) \\
& [-0.1112, -0.0633] \\
White Percent & -0.6496 ($p=0.0$) \\
& [-0.7052, -0.5939] \\
Total Housing Units & 0.5035 ($p=0.0$) \\
& [0.4499, 0.5571] \\
Median Household Income & -0.1162 ($p=0.0$) \\
& [-0.1325, -0.0999] \\
Unemployment Rate & 0.1059 ($p=0.0$) \\
& [0.0875, 0.1244] \\
Radius Of Gyration & -0.0357 ($p=0.0018$) \\
& [-0.058, -0.0133] \\
Entropy & 0.8344 ($p=0.0$) \\
& [0.8114, 0.8573] \\
Education Percent & -0.1705 ($p=0.0$) \\
& [-0.1925, -0.1485] \\

\hline \\[-1.8ex]
Observations & 86562 \\
Pseudo $R^2$ & 0.112 \\
\hline
\hline \\[-1.8ex]
\multicolumn{2}{r}{\textit{Note:} $^{*}p<0.1$; $^{**}p<0.05$; $^{***}p<0.01$} \\
\label{tab: Logistic_evac_480}
\end{tabular}
\end{table}

\begin{table}[!htbp] \centering
  \caption{Regression Results for Evacuation binary variable, Threshold: 720}
\begin{tabular}{@{\extracolsep{5pt}}lc}
\hline \hline
& \multicolumn{1}{c}{\textit{Dependent variable: Evacuation status}} \\
\cline{2-2}
\\[-1.8ex] & \textbf{Model 1} \\
\hline \\[-1.8ex]
Const & -2.0036 ($p=0.0$) \\
& [-2.0278, -1.9795] \\
Distance From Fire & -0.0508 ($p=0.0$) \\
& [-0.0718, -0.0299] \\
Population Density & -0.1087 ($p=0.0$) \\
& [-0.1319, -0.0854] \\
Asian Percent & 0.1373 ($p=0.0$) \\
& [0.1119, 0.1628] \\
Black Percent & -0.1136 ($p=0.0$) \\
& [-0.1395, -0.0877] \\
White Percent & -0.7671 ($p=0.0$) \\
& [-0.8282, -0.7061] \\
Total Housing Units & 0.6005 ($p=0.0$) \\
& [0.5419, 0.6592] \\
Median Household Income & -0.115 ($p=0.0$) \\
& [-0.1321, -0.0979] \\
Unemployment Rate & 0.1295 ($p=0.0$) \\
& [0.11, 0.149] \\
Radius Of Gyration & -0.0619 ($p=0.0$) \\
& [-0.0859, -0.0378] \\
Entropy & 0.8876 ($p=0.0$) \\
& [0.8628, 0.9125] \\
Education Percent & -0.1604 ($p=0.0$) \\
& [-0.1839, -0.137] \\

\hline \\[-1.8ex]
Observations & 77709 \\
Pseudo $R^2$ & 0.123 \\
\hline
\hline \\[-1.8ex]
\multicolumn{2}{r}{\textit{Note:} $^{*}p<0.1$; $^{**}p<0.05$; $^{***}p<0.01$} \\
\label{tab: Logistic_evac_720}
\end{tabular}
\end{table}

\begin{table}[!htbp] \centering
  \caption{Regression Results for Evacuation binary variable, Threshold: 960}
\begin{tabular}{@{\extracolsep{5pt}}lc}
\hline \hline
& \multicolumn{1}{c}{\textit{Dependent variable: Evacuation status}} \\
\cline{2-2}
\\[-1.8ex] & \textbf{Model 1} \\
\hline \\[-1.8ex]
Const & -2.0689 ($p=0.0$) \\
& [-2.0997, -2.0381] \\
Distance From Fire & -0.0463 ($p=0.0007$) \\
& [-0.0732, -0.0195] \\
Population Density & -0.1241 ($p=0.0$) \\
& [-0.1537, -0.0945] \\
Asian Percent & 0.1534 ($p=0.0$) \\
& [0.1215, 0.1853] \\
Black Percent & -0.1077 ($p=0.0$) \\
& [-0.1402, -0.0752] \\
White Percent & -0.8154 ($p=0.0$) \\
& [-0.8925, -0.7383] \\
Total Housing Units & 0.6185 ($p=0.0$) \\
& [0.5453, 0.6918] \\
Median Household Income & -0.0907 ($p=0.0$) \\
& [-0.1127, -0.0686] \\
Unemployment Rate & 0.1194 ($p=0.0$) \\
& [0.0948, 0.1441] \\
Radius Of Gyration & -0.0789 ($p=0.0$) \\
& [-0.11, -0.0478] \\
Entropy & 0.8837 ($p=0.0$) \\
& [0.8518, 0.9156] \\
Education Percent & -0.1819 ($p=0.0$) \\
& [-0.2112, -0.1526] \\

\hline \\[-1.8ex]
Observations & 50108 \\
Pseudo $R^2$ & 0.118 \\
\hline
\hline \\[-1.8ex]
\multicolumn{2}{r}{\textit{Note:} $^{*}p<0.1$; $^{**}p<0.05$; $^{***}p<0.01$} \\
\label{tab: Logistic_evac_960}
\end{tabular}
\end{table}

\begin{table}[!htbp] \centering
  \caption{Regression Results for Evacuation binary variable, Threshold: 1200}
\begin{tabular}{@{\extracolsep{5pt}}lc}
\hline \hline
& \multicolumn{1}{c}{\textit{Dependent variable: Evacuation status}} \\
\cline{2-2}
\\[-1.8ex] & \textbf{Model 1} \\
\hline \\[-1.8ex]
Const & -2.2695 ($p=0.0$) \\
& [-2.312, -2.2269] \\
Distance From Fire & -0.0395 ($p=0.0322$) \\
& [-0.0756, -0.0034] \\
Population Density & -0.1351 ($p=0.0$) \\
& [-0.1752, -0.0951] \\
Asian Percent & 0.143 ($p=0.0$) \\
& [0.1, 0.1859] \\
Black Percent & -0.084 ($p=0.0001$) \\
& [-0.1271, -0.041] \\
White Percent & -0.858 ($p=0.0$) \\
& [-0.9603, -0.7558] \\
Total Housing Units & 0.6404 ($p=0.0$) \\
& [0.5437, 0.7371] \\
Median Household Income & -0.0919 ($p=0.0$) \\
& [-0.1194, -0.0644] \\
Unemployment Rate & 0.1167 ($p=0.0$) \\
& [0.0842, 0.1492] \\
Radius Of Gyration & -0.079 ($p=0.0002$) \\
& [-0.1206, -0.0373] \\
Entropy & 0.9359 ($p=0.0$) \\
& [0.8929, 0.9789] \\
Education Percent & -0.1959 ($p=0.0$) \\
& [-0.2349, -0.157] \\

\hline \\[-1.8ex]
Observations & 31235 \\
Pseudo $R^2$ & 0.129 \\
\hline
\hline \\[-1.8ex]
\multicolumn{2}{r}{\textit{Note:} $^{*}p<0.1$; $^{**}p<0.05$; $^{***}p<0.01$} \\
\label{tab: Logistic_evac_1200}
\end{tabular}
\end{table}

\begin{table}[!htbp] \centering
  \caption{Regression Results for Evacuation binary variable, Threshold: 1440}
\begin{tabular}{@{\extracolsep{5pt}}lc}
\hline \hline
& \multicolumn{1}{c}{\textit{Dependent variable: Evacuation status}} \\
\cline{2-2}
\\[-1.8ex] & \textbf{Model 1} \\
\hline \\[-1.8ex]
Const & -2.6359 ($p=0.0$) \\
& [-2.8555, -2.4164] \\
Dist Epicentre & 0.0051 ($p=0.9562$) \\
& [-0.1767, 0.1869] \\
Population Density & -0.0307 ($p=0.7528$) \\
& [-0.222, 0.1605] \\
Asian Population & 0.2539 ($p=0.01$) \\
& [0.0606, 0.4471] \\
Black Population & -0.1469 ($p=0.1727$) \\
& [-0.3582, 0.0643] \\
White Population & -0.6894 ($p=0.0021$) \\
& [-1.1285, -0.2503] \\
Total Housing Units & 0.5797 ($p=0.0052$) \\
& [0.173, 0.9864] \\
Median Household Income & -0.1312 ($p=0.0278$) \\
& [-0.2481, -0.0143] \\
Unemployment Rate & 0.0385 ($p=0.654$) \\
& [-0.13, 0.2071] \\
Radius Of Gyration & 0.0922 ($p=0.3853$) \\
& [-0.1159, 0.3002] \\
Avg Entropy & 0.9655 ($p=0.0$) \\
& [0.7666, 1.1645] \\
Education Atleast One Degree & -0.1329 ($p=0.1793$) \\
& [-0.3267, 0.061] \\

\hline \\[-1.8ex]
Observations & 1677 \\
Pseudo $R^2$ & 0.150 \\
\hline
\hline \\[-1.8ex]
\multicolumn{2}{r}{\textit{Note:} $^{*}p<0.1$; $^{**}p<0.05$; $^{***}p<0.01$} \\
\label{tab: Logistic_evac_1440}
\end{tabular}
\end{table}

\begin{table}[!htbp] \centering
  \caption{Regression Results: Evacuation Distance regressed on Sociodemographic variables}
\begin{tabular}{@{\extracolsep{5pt}}lc}
\hline \hline
& \multicolumn{1}{c}{\textit{Dependent variable: Evacuation Distance}} \\
\cline{2-2}
\\[-1.8ex] & \textbf{Model 1} \\
\hline \\[-1.8ex]
Const & 4.2273 ($p=0.0$) \\
& [4.2157, 4.239] \\
Population Density & -0.0479 ($p=0.0$) \\
& [-0.0604, -0.0353] \\
Unemployment & 0.0161 ($p=0.0112$) \\
& [0.0037, 0.0285] \\
White Percent & 0.0056 ($p=0.5535$) \\
& [-0.0129, 0.024] \\
Black Percent & -0.0188 ($p=0.0346$) \\
& [-0.0362, -0.0014] \\
Asian Percent & -0.0159 ($p=0.0147$) \\
& [-0.0287, -0.0031] \\
Median Household Income & 0.052 ($p=0.0$) \\
& [0.0387, 0.0654] \\
Education Percent & -0.0268 ($p=0.0$) \\
& [-0.039, -0.0146] \\

\hline \\[-1.8ex]
Observations & 13560 \\
$R^2$ & 0.017 \\
Adjusted $R^2$ & 0.016 \\
\hline
\hline \\[-1.8ex]
\multicolumn{2}{r}{\textit{Note:} $^{*}p<0.1$; $^{**}p<0.05$; $^{***}p<0.01$} \\
\label{tab: ols_ED+SDM}
\end{tabular}
\end{table}
\begin{table}[!htbp] \centering
  \caption{Regression Results: Evacuation Distance regressed on sociodemographic variables and Distance from fire}
\begin{tabular}{@{\extracolsep{5pt}}lc}
\hline \hline
& \multicolumn{1}{c}{\textit{Dependent variable: Evacuation Distance}} \\
\cline{2-2}
\\[-1.8ex] & \textbf{Model 1} \\
\hline \\[-1.8ex]
Const & 4.2273 ($p=0.0$) \\
& [4.2157, 4.239] \\
Population Density & -0.0417 ($p=0.0$) \\
& [-0.0546, -0.0289] \\
Unemployment & 0.0169 ($p=0.0079$) \\
& [0.0044, 0.0293] \\
White Percent & 0.0045 ($p=0.6327$) \\
& [-0.0139, 0.0229] \\
Black Percent & -0.0212 ($p=0.0169$) \\
& [-0.0387, -0.0038] \\
Asian Percent & -0.0115 ($p=0.0813$) \\
& [-0.0244, 0.0014] \\
Median Household Income & 0.0519 ($p=0.0$) \\
& [0.0385, 0.0652] \\
Education Percent & -0.0242 ($p=0.0001$) \\
& [-0.0365, -0.0119] \\
Distance From Fire & 0.0294 ($p=0.0$) \\
& [0.0174, 0.0415] \\

\hline \\[-1.8ex]
Observations & 13560 \\
$R^2$ & 0.018 \\
Adjusted $R^2$ & 0.018 \\
\hline
\hline \\[-1.8ex]
\multicolumn{2}{r}{\textit{Note:} $^{*}p<0.1$; $^{**}p<0.05$; $^{***}p<0.01$} \\
\label{tab: ols_ED+SDM_fire}
\end{tabular}
\end{table}
\begin{table}[!htbp] \centering
  \caption{Regression Results: Evacuation Distance regressed with sociodemographic variables and distance from fire and pre disaster mobility}
\begin{tabular}{@{\extracolsep{5pt}}lc}
\hline \hline
& \multicolumn{1}{c}{\textit{Dependent variable: Evacuation Distance}} \\
\cline{2-2}
\\[-1.8ex] & \textbf{Model 1} \\
\hline \\[-1.8ex]
Const & 4.2273 ($p=0.0$) \\
& [4.2161, 4.2385] \\
Population Density & -0.0341 ($p=0.0$) \\
& [-0.0464, -0.0218] \\
Unemployment & 0.0119 ($p=0.0508$) \\
& [-0.0, 0.0238] \\
Radius Of Gyration & 0.2086 ($p=0.0$) \\
& [0.1967, 0.2205] \\
White Percent & -0.0048 ($p=0.5958$) \\
& [-0.0225, 0.0129] \\
Black Percent & -0.0143 ($p=0.0927$) \\
& [-0.031, 0.0024] \\
Asian Percent & -0.0098 ($p=0.1209$) \\
& [-0.0222, 0.0026] \\
Entropy & -0.0841 ($p=0.0$) \\
& [-0.0959, -0.0723] \\
Median Household Income & 0.0441 ($p=0.0$) \\
& [0.0312, 0.0569] \\
Education Percent & -0.0148 ($p=0.0139$) \\
& [-0.0265, -0.003] \\
Distance From Fire & 0.0158 ($p=0.0076$) \\
& [0.0042, 0.0275] \\

\hline \\[-1.8ex]
Observations & 13560 \\
$R^2$ & 0.098 \\
Adjusted $R^2$ & 0.098 \\
\hline
\hline \\[-1.8ex]
\multicolumn{2}{r}{\textit{Note:} $^{*}p<0.1$; $^{**}p<0.05$; $^{***}p<0.01$} \\
\label{tab: ols_ED+SDM_fire+pdm}
\end{tabular}
\end{table}

\begin{table}[!htbp] \centering
  \caption{Regression Results: Connectedness regressed with evacuation distance and median income}
\begin{tabular}{@{\extracolsep{5pt}}lc}
\hline \hline
& \multicolumn{1}{c}{\textit{Dependent variable: Connectedness}} \\
\cline{2-2}
\\[-1.8ex] & \textbf{Model 1} \\
\hline \\[-1.8ex]
Const & 13.7979 ($p=0.0$) \\
& [13.7839, 13.8118] \\
Median Household Income & -0.042 ($p=0.0$) \\
& [-0.056, -0.0281] \\
Evacuation Distance & -0.8104 ($p=0.0$) \\
& [-0.8243, -0.7964] \\

\hline \\[-1.8ex]
Observations & 13560 \\
$R^2$ & 0.494 \\
Adjusted $R^2$ & 0.493 \\
\hline
\hline \\[-1.8ex]
\multicolumn{2}{r}{\textit{Note:} $^{*}p<0.1$; $^{**}p<0.05$; $^{***}p<0.01$} \\
\label{tab: Table S9}
\end{tabular}
\end{table}
\begin{table}[!htbp] \centering
  \caption{Regression Results: Connectedness regressed with evacuation Distance $+$ White Percent}
\begin{tabular}{@{\extracolsep{5pt}}lc}
\hline \hline
& \multicolumn{1}{c}{\textit{Dependent variable: Connectedness}} \\
\cline{2-2}
\\[-1.8ex] & \textbf{Model 1} \\
\hline \\[-1.8ex]
Const & 13.7979 ($p=0.0$) \\
& [13.7841, 13.8116] \\
White Percent & 0.1231 ($p=0.0$) \\
& [0.1093, 0.1368] \\
Evacuation Distance & -0.8217 ($p=0.0$) \\
& [-0.8355, -0.8079] \\

\hline \\[-1.8ex]
Observations & 13560 \\
$R^2$ & 0.503 \\
Adjusted $R^2$ & 0.503 \\
\hline
\hline \\[-1.8ex]
\multicolumn{2}{r}{\textit{Note:} $^{*}p<0.1$; $^{**}p<0.05$; $^{***}p<0.01$} \\
\label{tab: Table S10}
\end{tabular}
\end{table}
\begin{table}[!htbp] \centering
  \caption{Regression Results: Connectedness regressed with Evacuation Distance and Black Percent}
\begin{tabular}{@{\extracolsep{5pt}}lc}
\hline \hline
& \multicolumn{1}{c}{\textit{Dependent variable: Connectedness}} \\
\cline{2-2}
\\[-1.8ex] & \textbf{Model 1} \\
\hline \\[-1.8ex]
Const & 13.7979 ($p=0.0$) \\
& [13.7841, 13.8116] \\
Black Percent & -0.1337 ($p=0.0$) \\
& [-0.1475, -0.12] \\
Evacuation Distance & -0.8229 ($p=0.0$) \\
& [-0.8366, -0.8091] \\

\hline \\[-1.8ex]
Observations & 13560 \\
$R^2$ & 0.505 \\
Adjusted $R^2$ & 0.505 \\
\hline
\hline \\[-1.8ex]
\multicolumn{2}{r}{\textit{Note:} $^{*}p<0.1$; $^{**}p<0.05$; $^{***}p<0.01$} \\
\label{tab: Table S11}
\end{tabular}
\end{table}
\begin{table}[!htbp] \centering
  \caption{Regression Results: Similarity regressed with Evacuation Distance and Median Income}
\begin{tabular}{@{\extracolsep{5pt}}lc}
\hline \hline
& \multicolumn{1}{c}{\textit{Dependent variable: Similarity}} \\
\cline{2-2}
\\[-1.8ex] & \textbf{Model 1} \\
\hline \\[-1.8ex]
Const & 0.1816 ($p=0.0$) \\
& [0.174, 0.1891] \\
Median Income & 0.0339 ($p=0.0$) \\
& [0.0263, 0.0415] \\
Evacuation Distance & -0.1195 ($p=0.0$) \\
& [-0.1271, -0.1119] \\

\hline \\[-1.8ex]
Observations & 13560 \\
$R^2$ & 0.068 \\
Adjusted $R^2$ & 0.068 \\
\hline
\hline \\[-1.8ex]
\multicolumn{2}{r}{\textit{Note:} $^{*}p<0.1$; $^{**}p<0.05$; $^{***}p<0.01$} \\
\label{tab: Table S12}
\end{tabular}
\end{table}
\begin{table}[!htbp] \centering
  \caption{Regression Results: Similarity regressed with Evacuation Distance and White percent}
\begin{tabular}{@{\extracolsep{5pt}}lc}
\hline \hline
& \multicolumn{1}{c}{\textit{Dependent variable: Similarity}} \\
\cline{2-2}
\\[-1.8ex] & \textbf{Model 1} \\
\hline \\[-1.8ex]
Const & 0.1816 ($p=0.0$) \\
& [0.174, 0.1891] \\
White Percent & 0.006 ($p=0.1192$) \\
& [-0.0016, 0.0136] \\
Evacuation Distance & -0.1168 ($p=0.0$) \\
& [-0.1244, -0.1092] \\

\hline \\[-1.8ex]
Observations & 13560 \\
$R^2$ & 0.063 \\
Adjusted $R^2$ & 0.063 \\
\hline
\hline \\[-1.8ex]
\multicolumn{2}{r}{\textit{Note:} $^{*}p<0.1$; $^{**}p<0.05$; $^{***}p<0.01$} \\
\label{tab: Table S13}
\end{tabular}
\end{table}
\begin{table}[!htbp] \centering
  \caption{Regression Results: Similarity regressed with Evacuation Distance and Black percent}
\begin{tabular}{@{\extracolsep{5pt}}lc}
\hline \hline
& \multicolumn{1}{c}{\textit{Dependent variable: Similarity}} \\
\cline{2-2}
\\[-1.8ex] & \textbf{Model 1} \\
\hline \\[-1.8ex]
Const & 0.1816 ($p=0.0$) \\
& [0.174, 0.1891] \\
Black Percent & -0.0172 ($p=0.0$) \\
& [-0.0248, -0.0096] \\
Evacuation Distance & -0.1175 ($p=0.0$) \\
& [-0.1251, -0.11] \\

\hline \\[-1.8ex]
Observations & 13560 \\
$R^2$ & 0.064 \\
Adjusted $R^2$ & 0.064 \\
\hline
\hline \\[-1.8ex]
\multicolumn{2}{r}{\textit{Note:} $^{*}p<0.1$; $^{**}p<0.05$; $^{***}p<0.01$} \\
\label{tab: Table S14}
\end{tabular}
\end{table}

\begin{table}[!htbp] \centering
  \caption{Regression Results: Change in connectedness regressed with Evacuation Distance and Median income}
\begin{tabular}{@{\extracolsep{5pt}}lc}
\hline \hline
& \multicolumn{1}{c}{\textit{Dependent variable: $\delta$ Connectedness}} \\
\cline{2-2}
\\[-1.8ex] & \textbf{Model 1} \\
\hline \\[-1.8ex]
Const & 1.3873 ($p=0.0$) \\
& [1.3674, 1.4072] \\
Median Income & -0.0521 ($p=0.0$) \\
& [-0.0721, -0.0321] \\
Log Distance & -0.8337 ($p=0.0$) \\
& [-0.8538, -0.8137] \\

\hline \\[-1.8ex]
Observations & 15962 \\
$R^2$ & 0.301 \\
Adjusted $R^2$ & 0.301 \\
\hline
\hline \\[-1.8ex]
\multicolumn{2}{r}{\textit{Note:} $^{*}p<0.1$; $^{**}p<0.05$; $^{***}p<0.01$} \\
\label{tab: Table S15}
\end{tabular}
\end{table}
\begin{table}[!htbp] \centering
  \caption{Regression Results: Change in connectedness regressed with Evacuation Distance and White percent}
\begin{tabular}{@{\extracolsep{5pt}}lc}
\hline \hline
& \multicolumn{1}{c}{\textit{Dependent variable: $\delta$ Connectedness}} \\
\cline{2-2}
\\[-1.8ex] & \textbf{Model 1} \\
\hline \\[-1.8ex]
Const & 1.3873 ($p=0.0$) \\
& [1.3675, 1.4071] \\
White Percent & 0.1516 ($p=0.0$) \\
& [0.1318, 0.1714] \\
Log Distance & -0.8504 ($p=0.0$) \\
& [-0.8702, -0.8306] \\

\hline \\[-1.8ex]
Observations & 15962 \\
$R^2$ & 0.310 \\
Adjusted $R^2$ & 0.310 \\
\hline
\hline \\[-1.8ex]
\multicolumn{2}{r}{\textit{Note:} $^{*}p<0.1$; $^{**}p<0.05$; $^{***}p<0.01$} \\
\label{tab: Table S16}
\end{tabular}
\end{table}

\begin{table}[!htbp] \centering
  \caption{Regression Results: Change in connectedness regressed with Evacuation Distance and Black percent}
\begin{tabular}{@{\extracolsep{5pt}}lc}
\hline \hline
& \multicolumn{1}{c}{\textit{Dependent variable: $\delta$ Connectedness}} \\
\cline{2-2}
\\[-1.8ex] & \textbf{Model 1} \\
\hline \\[-1.8ex]
Const & 1.3873 ($p=0.0$) \\
& [1.3676, 1.407] \\
Black Percent & -0.1834 ($p=0.0$) \\
& [-0.2032, -0.1637] \\
Log Distance & -0.8557 ($p=0.0$) \\
& [-0.8754, -0.8359] \\

\hline \\[-1.8ex]
Observations & 15962 \\
$R^2$ & 0.314 \\
Adjusted $R^2$ & 0.314 \\
\hline
\hline \\[-1.8ex]
\multicolumn{2}{r}{\textit{Note:} $^{*}p<0.1$; $^{**}p<0.05$; $^{***}p<0.01$} \\
\label{tab: Table S17}
\end{tabular}
\end{table}
\begin{table}[!htbp] \centering
  \caption{Regression Results: Change in similarity regressed with Evacuation Distance and Median income}
\begin{tabular}{@{\extracolsep{5pt}}lc}
\hline \hline
& \multicolumn{1}{c}{\textit{Dependent variable: $\delta$ Similarity}} \\
\cline{2-2}
\\[-1.8ex] & \textbf{Model 1} \\
\hline \\[-1.8ex]
Const & -0.1631 ($p=0.0$) \\
& [-0.1731, -0.1531] \\
Median Income & 0.0053 ($p=0.3014$) \\
& [-0.0048, 0.0154] \\
Log Distance & 0.1302 ($p=0.0$) \\
& [0.1202, 0.1403] \\

\hline \\[-1.8ex]
Observations & 15962 \\
$R^2$ & 0.040 \\
Adjusted $R^2$ & 0.039 \\
\hline
\hline \\[-1.8ex]
\multicolumn{2}{r}{\textit{Note:} $^{*}p<0.1$; $^{**}p<0.05$; $^{***}p<0.01$} \\
\label{tab: Table S18}
\end{tabular}
\end{table}
\begin{table}[!htbp] \centering
  \caption{Regression Results: Change in similarity regressed with Evacuation Distance and White percent}
\begin{tabular}{@{\extracolsep{5pt}}lc}
\hline \hline
& \multicolumn{1}{c}{\textit{Dependent variable: $\delta$ Similarity}} \\
\cline{2-2}
\\[-1.8ex] & \textbf{Model 1} \\
\hline \\[-1.8ex]
Const & -0.1631 ($p=0.0$) \\
& [-0.1731, -0.1531] \\
White Percent & 0.038 ($p=0.0$) \\
& [0.028, 0.048] \\
Log Distance & 0.1282 ($p=0.0$) \\
& [0.1181, 0.1382] \\

\hline \\[-1.8ex]
Observations & 15962 \\
$R^2$ & 0.043 \\
Adjusted $R^2$ & 0.043 \\
\hline
\hline \\[-1.8ex]
\multicolumn{2}{r}{\textit{Note:} $^{*}p<0.1$; $^{**}p<0.05$; $^{***}p<0.01$} \\
\label{tab: Table S19}
\end{tabular}
\end{table}
\begin{table}[!htbp] \centering
  \caption{Regression Results: Change in Similarity regressed with evacuation distance and Black percent}
\begin{tabular}{@{\extracolsep{5pt}}lc}
\hline \hline
& \multicolumn{1}{c}{\textit{Dependent variable: $\delta$ Similarity}} \\
\cline{2-2}
\\[-1.8ex] & \textbf{Model 1} \\
\hline \\[-1.8ex]
Const & -0.1631 ($p=0.0$) \\
& [-0.1731, -0.1531] \\
Black Percent & -0.0344 ($p=0.0$) \\
& [-0.0445, -0.0244] \\
Log Distance & 0.1278 ($p=0.0$) \\
& [0.1178, 0.1379] \\

\hline \\[-1.8ex]
Observations & 15962 \\
$R^2$ & 0.042 \\
Adjusted $R^2$ & 0.042 \\
\hline
\hline \\[-1.8ex]
\multicolumn{2}{r}{\textit{Note:} $^{*}p<0.1$; $^{**}p<0.05$; $^{***}p<0.01$} \\
\label{tab: Table S20}
\end{tabular}
\end{table}
\begin{table}[!htbp] \centering
  \caption{Week 1 (Directly Impacted- Y=1, N=0): Predictor - Connectedness}
\begin{tabular}{@{\extracolsep{5pt}}lccc}
\hline \hline
& \textbf{Coefficient} & \textbf{P-value} & \textbf{95\% Confidence Interval} \\
\hline
Intercept & -0.0741 & 0.4961 & [-0.2875, 0.1393] \\
Connectedness & 0.0001 & 0.9994 & [-0.2525, 0.2527] \\

\hline
Pseudo $R^2$ & 0.000 & Observations & 351 \\
\hline \hline
\textit{Note:} & \multicolumn{3}{r}{$^{*}$p$<$0.1; $^{**}$p$<$0.05; $^{***}$p$<$0.01} \\
\label{tab: Table S21}
\end{tabular}
\end{table}
\begin{table}[!htbp] \centering
  \caption{Week 1 (Directly Impacted - Y=1, N=0): Predictor - Similarity}
\begin{tabular}{@{\extracolsep{5pt}}lccc}
\hline \hline
& \textbf{Coefficient} & \textbf{P-value} & \textbf{95\% Confidence Interval} \\
\hline
Intercept & -0.0626 & 0.5607 & [-0.2733, 0.1482] \\
Similarity & 0.1888$^{*}$ & 0.058 & [-0.0064, 0.384] \\

\hline
Pseudo $R^2$ & 0.007 & Observations & 351 \\
\hline \hline
\textit{Note:} & \multicolumn{3}{r}{$^{*}$p$<$0.1; $^{**}$p$<$0.05; $^{***}$p$<$0.01} \\
\label{tab: Table S22}
\end{tabular}
\end{table}
\begin{table}[!htbp] \centering
  \caption{Week 20 (Directly Impacted- Y=1, N=0): Predictor - Connectedness}
\begin{tabular}{@{\extracolsep{5pt}}lccc}
\hline \hline
& \textbf{Coefficient} & \textbf{P-value} & \textbf{95\% Confidence Interval} \\
\hline
Intercept & -1.044 & 0.0 & [-1.2927, -0.7954] \\
Connectedness & 0.3839 & 0.0045 & [0.1192, 0.6485] \\

\hline
Pseudo $R^2$ & 0.021 & Observations & 351 \\
\hline \hline
\textit{Note:} & \multicolumn{3}{r}{$^{*}$p$<$0.1; $^{**}$p$<$0.05; $^{***}$p$<$0.01} \\
\label{tab: Table S23}
\end{tabular}
\end{table}
\begin{table}[!htbp] \centering
  \caption{Week 20 (Directly Impacted- Y=1, N=0): Predictor - Similarity}
\begin{tabular}{@{\extracolsep{5pt}}lccc}
\hline \hline
& \textbf{Coefficient} & \textbf{P-value} & \textbf{95\% Confidence Interval} \\
\hline
Intercept & -1.125 & 0.0 & [-1.3687, -0.8814] \\
Similarity & 0.0053 & 0.963 & [-0.2194, 0.23] \\

\hline
Pseudo $R^2$ & 0.000 & Observations & 351 \\
\hline \hline
\textit{Note:} & \multicolumn{3}{r}{$^{*}$p$<$0.1; $^{**}$p$<$0.05; $^{***}$p$<$0.01} \\
\label{tab: Table S24}
\end{tabular}
\end{table}
\begin{table}[!htbp] \centering
  \caption{Week 1 (Precautionary- Y=1, N=0): Predictor - Similarity}
\begin{tabular}{@{\extracolsep{5pt}}lccc}
\hline \hline
& \textbf{Coefficient} & \textbf{P-value} & \textbf{95\% Confidence Interval} \\
\hline
Intercept & 0.3649 & 0.0 & [0.3331, 0.3966] \\
Similarity & -0.0206 & 0.2035 & [-0.0525, 0.0112] \\

\hline
Pseudo $R^2$ & 0.000 & Observations & 15764 \\
\hline \hline
\textit{Note:} & \multicolumn{3}{r}{$^{*}$p$<$0.1; $^{**}$p$<$0.05; $^{***}$p$<$0.01} \\
\label{tab: Table S25}
\end{tabular}
\end{table}
\begin{table}[!htbp] \centering
  \caption{Week 1 (Precautionary- Y=1, N=0): Predictor - Connectedness}
\begin{tabular}{@{\extracolsep{5pt}}lccc}
\hline \hline
& \textbf{Coefficient} & \textbf{P-value} & \textbf{95\% Confidence Interval} \\
\hline
Intercept & -1.3152 & 0.0 & [-1.3537, -1.2767] \\
Connectedness & 0.1865 & 0.0 & [0.1487, 0.2243] \\

\hline
Pseudo $R^2$ & 0.006 & Observations & 15720 \\
\hline \hline
\textit{Note:} & \multicolumn{3}{r}{$^{*}$p$<$0.1; $^{**}$p$<$0.05; $^{***}$p$<$0.01} \\
\label{tab: Table S26}
\end{tabular}
\end{table}
\begin{table}[!htbp] \centering
  \caption{Week 1 (Precautionary- Y=1, N=0): Predictor - Similarity}
\begin{tabular}{@{\extracolsep{5pt}}lccc}
\hline \hline
& \textbf{Coefficient} & \textbf{P-value} & \textbf{95\% Confidence Interval} \\
\hline
Intercept & 0.3649 & 0.0 & [0.3331, 0.3966] \\
Similarity & -0.0206 & 0.2035 & [-0.0525, 0.0112] \\

\hline
Pseudo $R^2$ & 0.000 & Observations & 15764 \\
\hline \hline
\textit{Note:} & \multicolumn{3}{r}{$^{*}$p$<$0.1; $^{**}$p$<$0.05; $^{***}$p$<$0.01} \\
\label{tab: Table S27}
\end{tabular}
\end{table}
\begin{table}[!htbp] \centering
  \caption{Week 20 (Precautionary- Y=1, N=0): Predictor - Similarity}
\begin{tabular}{@{\extracolsep{5pt}}lccc}
\hline \hline
& \textbf{Coefficient} & \textbf{P-value} & \textbf{95\% Confidence Interval} \\
\hline
Intercept & -1.3079 & 0.0 & [-1.3462, -1.2696] \\
Similarity & -0.173 & 0.0 & [-0.211, -0.135] \\

\hline
Pseudo $R^2$ & 0.005 & Observations & 15764 \\
\hline \hline
\textit{Note:} & \multicolumn{3}{r}{$^{*}$p$<$0.1; $^{**}$p$<$0.05; $^{***}$p$<$0.01} \\
\label{tab: Table S28}
\end{tabular}
\end{table}

\bibliographystyle{plain}
\bibliography{sample}